\documentclass[11pt,twoside,a4paper,cmspaper,final,collab]{cms-tdr}
\def\svnVersion{00d4abe-D}\def\svnDate{2026/09/01}\def\cmsCernNoTag{CERN-EP-2026-197}\def\cmsCernDate{\today}\def\cmsMessage{Submitted to Physics Letters B}
\begin{document}\cmsNoteHeader{HIG-25-010}

\newcommand{\cmsTable}[1]{\resizebox{\textwidth}{!}{#1}}
\ifthenelse{\boolean{cms@external}}
{
  \newcommand{\cmsTableIf}[1]{#1}
}
{
    \newcommand{\cmsTableIf}[1]{\resizebox{\textwidth}{!}{#1}}
}
\newlength\cmsTabSkip\setlength\cmsTabSkip{1ex}

\newcommand{\MiNNLO}{\textsc{MiNNLO}$_\mathrm{PS}$\xspace}

\newcommand{\Zll}{\ensuremath{\PZ \to \Pellp \Pellm}\xspace}
\newcommand{\Zllg}{\ensuremath{\PZ \to \Pellp \Pellm \PGg}\xspace}
\newcommand{\Zee}{\ensuremath{\PZ \to \Pep \Pem}\xspace}
\newcommand{\Zmm}{\ensuremath{\PZ \to \PGmp \PGmm}\xspace}
\newcommand{\Zmmg}{\ensuremath{\PZ \to \PGmp \PGmm \PGg}\xspace}
\newcommand{\gjets}{\ensuremath{\PGg+\text{jets}}\xspace}
\newcommand{\smzg}{\ensuremath{\PZ+\PGg}\xspace}
\newcommand{\zfake}{\ensuremath{\PZ+\text{fake photon}}\xspace}
\newcommand{\tttw}{\ensuremath{\PQt\PAQt/\PQt\PW}\xspace}
\newcommand{\ttX}{\ensuremath{\PQt\PAQt+\PW/\PZ}\xspace}
\newcommand{\tgttg}{\ensuremath{\PQt\PGg/\PQt\PAQt\PGg}\xspace}
\newcommand{\vbszg}{\ensuremath{\text{VBS }\PZ(+\PGg)}\xspace}
\newcommand{\mmg}{\ensuremath{\Pgmp \Pgmm \Pgg}\xspace}
\newcommand{\eeg}{\ensuremath{\Pep \Pem \Pgg}\xspace}
\newcommand{\Hmm}{\ensuremath{\PH \to \PGmp \PGmm}\xspace}
\newcommand{\Hmmgfsr}{\ensuremath{\Hmm+\PGg_{\mathrm{FSR}}}\xspace}
\newcommand{\HZG}{\ensuremath{\PH \to \PZ \Pgg}\xspace}
\newcommand{\ttH}{\ensuremath{\ttbar\PH}\xspace}
\newcommand{\VH}{\ensuremath{\PV\PH}\xspace}
\newcommand{\WH}{\ensuremath{\PWpm\PH}\xspace}
\newcommand{\ZH}{\ensuremath{\PZ\PH}\xspace}
\newcommand{\ppH}{\ensuremath{\Pp\Pp \to \PH}\xspace}
\newcommand{\llg}{\ensuremath{\Pellp \Pellm \Pgg}\xspace}

\newcommand{\mH}{\ensuremath{m_{\PH}}\xspace}
\newcommand{\mq}{\ensuremath{m_\mathrm{q}}\xspace}

\newcommand{\tspace}{\rule{0pt}{3ex}}
\newcommand{\bspace}{\rule[-2.0ex]{0pt}{0pt}}

\newcommand{\ptl}{\ensuremath{\pt(\Pell)}\xspace}
\newcommand{\nl}{\ensuremath{N_{\Pell}}\xspace}
\newcommand{\njet}{\ensuremath{N_\text{jet}}\xspace}
\newcommand{\mllg}{\ensuremath{m_{\Pell \Pell \Pgg}}\xspace}
\newcommand{\mll}{\ensuremath{m_{\Pell \Pell}}\xspace}
\newcommand{\ptg}{\ensuremath{\pt(\Pgg)}\xspace}
\newcommand{\nbm}{\ensuremath{N_{\PQb}}\xspace}
\newcommand{\ptunderllg}{\ensuremath{p_{\mathrm{T},\Pell \Pell \Pgg}}\xspace}
\newcommand{\stmiss}{\ensuremath{\vec{S}_\mathrm{T}^\text{miss}}\xspace}
\newcommand{\deltar}{\ensuremath{\Delta R}\xspace}
\newcommand{\drmin}{\ensuremath{\Delta R_\text{min}(\PGg, \Pell)}\xspace}
\newcommand{\drmax}{\ensuremath{\Delta R_\text{max}(\PGg, \Pell)}\xspace}
\newcommand{\dphiphstmiss}{\tspace\bspace\ensuremath{\Delta\phi(\PGg,\stmiss)}\xspace}
\newcommand{\Irel}{\ensuremath{I_{\text{rel}}}\xspace}
\newcommand{\Imini}{\ensuremath{I_{\text{mini}}}\xspace}
\newcommand{\etag}{\ensuremath{\eta(\Pgg)}\xspace}
\newcommand{\etalo}{\ensuremath{\eta(\Pell_1)}\xspace}
\newcommand{\etalt}{\ensuremath{\eta(\Pell_2)}\xspace}
\newcommand{\mjj}{\ensuremath{m_{\mathrm{jj}}}\xspace}
\newcommand{\sysbalance}{\tspace\bspace\ensuremath{\abs{\sum_{\Pell\Pell,\PGg,\text{jets}} \vec{p}_\mathrm{T}^{i}}/\sum_{\Pell\Pell,\PGg,\text{jets}} p_\mathrm{T}^{i}}\xspace}
\newcommand{\zeppenfeld}{\tspace\bspace\ensuremath{\abs{\eta_{\PGg}-\frac{\eta_{j_1}+\eta_{j_2}}{2}}}\xspace}
\newcommand{\detajj}{\ensuremath{\Delta \eta(\text{j}_{1},\text{j}_{2})}\xspace}
\newcommand{\dphillgj}{\ensuremath{\Delta\phi(\llg,\text{j})}\xspace}
\newcommand{\dphijj}{\ensuremath{\Delta \phi(\text{j}_{1},\text{j}_{2})}\xspace}
\newcommand{\dphillgjj}{\ensuremath{\Delta \phi(\llg,\text{jj})}\xspace}

\newcommand{\vhthreelep}{\ensuremath{\VH\, N_{\Pell} \geq 3}\xspace}
\newcommand{\vhptmiss}{\ensuremath{\VH} \ptmiss\xspace}
\newcommand{\tthlep}{\ensuremath{\ttH} leptonic\xspace}
\newcommand{\tthhad}{\ensuremath{\ttH} hadronic\xspace}

\newcommand{\sigmarel}{\ensuremath{\sigma_E(\PGg)/E(\PGg)}\xspace}
\newcommand{\sigmaeff}{\ensuremath{\sigma_{\text{eff}}}\xspace}
\newcommand{\expsig}{\ensuremath{Z_{\mathrm{cc}}}\xspace}
\newcommand{\nsig}{\ensuremath{N_{\mathrm{S}}}\xspace}
\newcommand{\nbkg}{\ensuremath{N_{\mathrm{B}}}\xspace}
\newcommand{\zetas}{\ensuremath{\zeta_s}\xspace}
\newcommand{\deltamc}{\ensuremath{\Delta_{\mathrm{MC}}}\xspace}
\newcommand{\nspurious}{\ensuremath{N_{\mathrm{s}}}\xspace}

\cmsNoteHeader{HIG-25-010}
\title{Search for the rare Higgs boson decay \texorpdfstring{$\PH\to\PZ\PGg$}{H to Z gamma} in proton-proton collisions at \texorpdfstring{$\sqrt{s}=13$}{sqrt(s) = 13} and \texorpdfstring{$13.6\TeV$}{13.6 TeV}}

\date{\today}

\abstract{A search is presented for the rare, loop-induced Higgs boson decay $\PH\to\PZ\PGg$, where $\PZ\to \Pellp\Pellm$ and $\Pell = \Pe$ or $\PGm$. The search is performed using a sample of proton-proton ($\Pp\Pp$) collision data at the center-of-mass energies of 13 and 13.6\TeV, recorded by the CMS experiment at the LHC and corresponding to a total integrated luminosity of 200\fbinv. The analysis design separately considers and optimizes sensitivity to Higgs boson production via gluon-gluon fusion, vector boson fusion, and associated-production processes. The signal is extracted from a simultaneous fit to the invariant mass distributions of the $\Pellp\Pellm\PGg$ system in the various production channels and event categories. The observed (expected) signal strength $\mu$, defined as the ratio of the measured product the of cross section and decay branching fraction value $\sigma(\Pp\Pp\to\PH)\mathcal{B}(\PH\to\PZ\PGg)$ to the corresponding value predicted in the standard model, is found to be $\mu = 1.10^{+0.52}_{-0.61}\ (1.00^{+0.49}_{-0.46})$ for a Higgs boson mass of 125.38\GeV. The signal has an observed (expected) significance of 1.9 (2.3) standard deviations. }

\hypersetup{%
pdfauthor={CMS Collaboration},%
pdftitle={Search for the rare Higgs boson decay H to Z gamma in proton-proton collisions at sqrt(s) = 13 and 13.6 TeV},%
pdfsubject={CMS},%
pdfkeywords={CMS, Higgs boson}}

\maketitle 

\section{Introduction\label{Sec-Introduction}}

The discovery of the Higgs boson (\PH) in 2012 by the ATLAS and CMS
Collaborations~\cite{ATLAS:2012yve,CMS:2012qbp,CMS:2013btf}, together with subsequent measurements of its properties,
provides strong evidence that the spontaneous breaking of electroweak symmetry is responsible for the
generation of the masses of the fundamental particles of the standard model (SM), with the possible exception of neutrinos.
Following this discovery, ATLAS and CMS initiated an extensive program of measurements~\cite{ATLAS:2016neq,ATLAS:2022vkf,CMS:2022dwd} to test SM predictions across as many Higgs boson production mechanisms and decay modes as possible. When measured with sufficient precision, these observables can be sensitive to contributions from non-SM amplitudes, thereby providing an important probe of hypothetical scenarios for physics beyond the SM (BSM). To date, the measurements indicate that the values of the Higgs boson couplings and related properties are consistent with the SM predictions when the measured Higgs boson mass is used as an input to theoretical calculations. This program of measurements is far from complete, however, as several processes have not yet been observed and essentially all measurements would benefit from larger data samples.  

The process $\HZG$~\cite{Cahn:1978nz,Bergstrom:1985hp,Spira:1991tj}
is a relatively rare decay, with a predicted SM branching fraction
$\mathcal{B}(\HZG) = (1.5 \pm 0.1) \times 10^{-3}$~\cite{Djouadi:1997yw,deFlorian:2016spz}.
The dominant SM Feynman diagrams for $\HZG$ are shown in Fig.~\ref{fig:feynman}. Like the related process $\PH\to\PGg\PGg$,
the decay $\HZG$ proceeds in the SM only through loop-induced amplitudes, in contrast to $\PH\to\PZ\PZ^{*}$, where $\PZ^{*}$ is an off-mass-shell \PZ boson.  
The dominant SM contributions arise from intermediate states involving $\PW$ bosons and, to a lesser extent, top quarks.
This channel is therefore sensitive to additional contributions that could arise if BSM particles enter these loop diagrams~\cite{Low:2011gn,Low:2012rj,Chiang:2012qz,Carena:2012xa,Chen:2013vi,Azatov:2013ura,Dawson:2018pyl,Cao:2018cms}. 
Recent measurements of $\HZG$, discussed below,
have motivated further theoretical investigations, aimed both
at refining the SM predictions and at exploring possible effects of BSM physics~\cite{Boto:2023bpg,He:2025kzk,ReyesR:2025dok,Kachanovich:2025cxz,Mantzaropoulos:2024vpe,Israr:2024ubp,Sang:2024vqk,Hernandez-Juarez:2024iwe,Chen:2024vyn,He:2024bxi,Cheung:2024kml}. The $\HZG$ process has been found to be 
particularly relevant for probing the possibility that the Higgs boson is a composite Nambu--Goldstone
boson arising from a strongly interacting sector~\cite{Azatov:2013ura, Cao:2018cms}.

\ifthenelse{\boolean{cms@external}}
{
\begin{figure*}[tb!]
  \centering
  \includegraphics[width=0.3\textwidth]{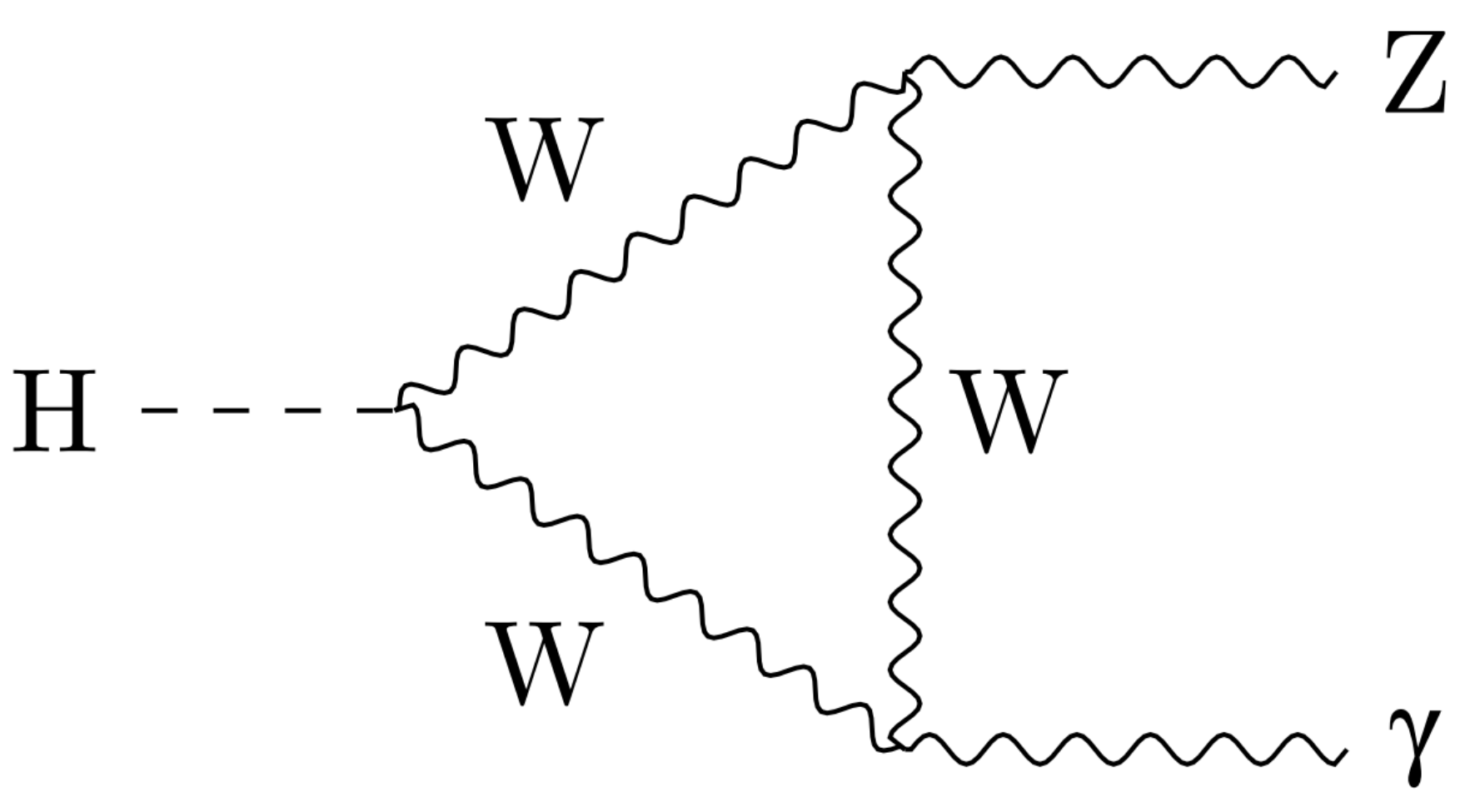}
  \includegraphics[width=0.3\textwidth]{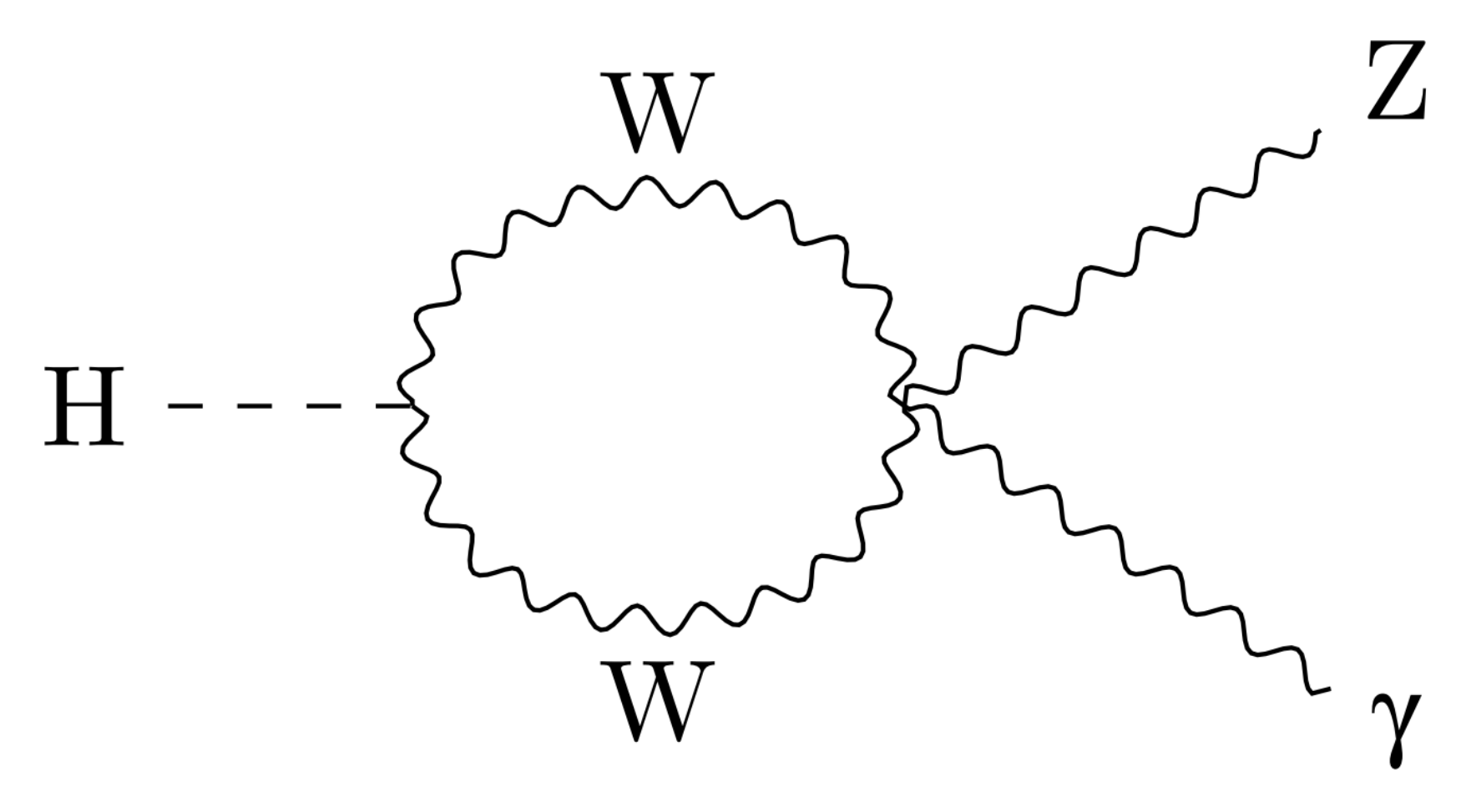}
  \includegraphics[width=0.3\textwidth]{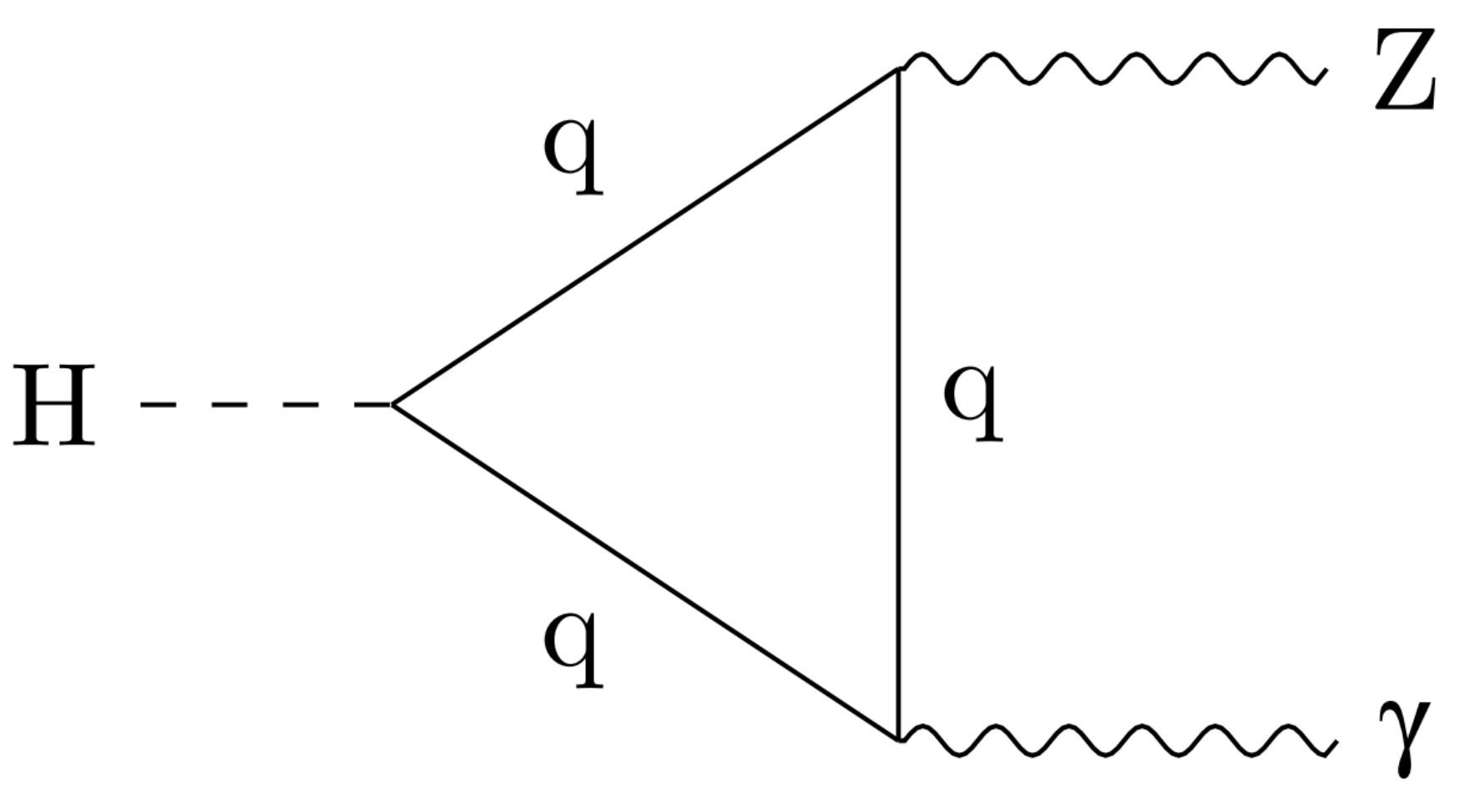}
  \caption{Feynman diagrams for $\HZG$ in the SM. In these loop-induced processes, the amplitudes associated with $\PW$ bosons in the
  intermediate state dominate over those with quarks.}
\label{fig:feynman}
\end{figure*}
}
{
\begin{figure*}[b!]
  \centering
  \includegraphics[width=0.3\textwidth]{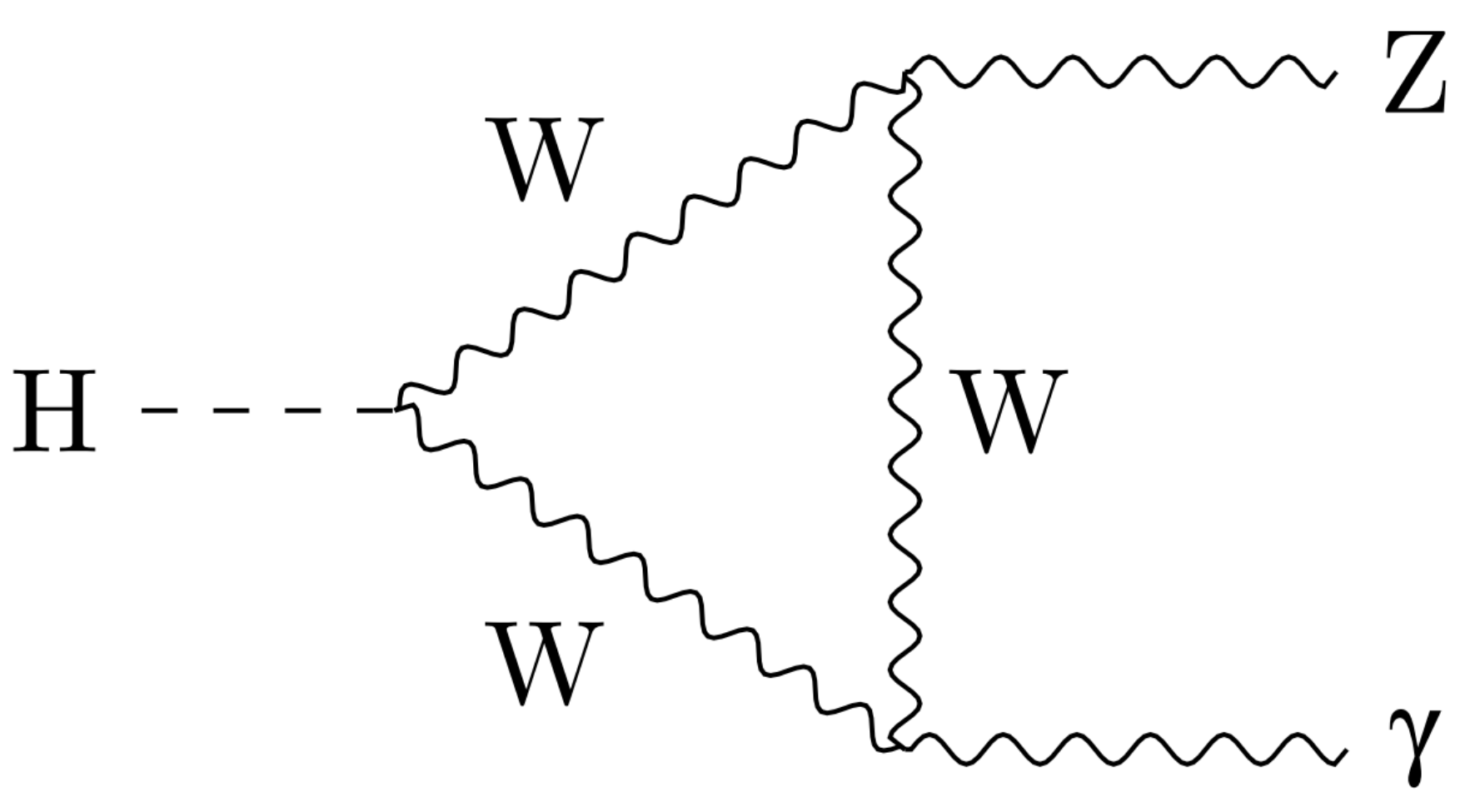}
  \includegraphics[width=0.3\textwidth]{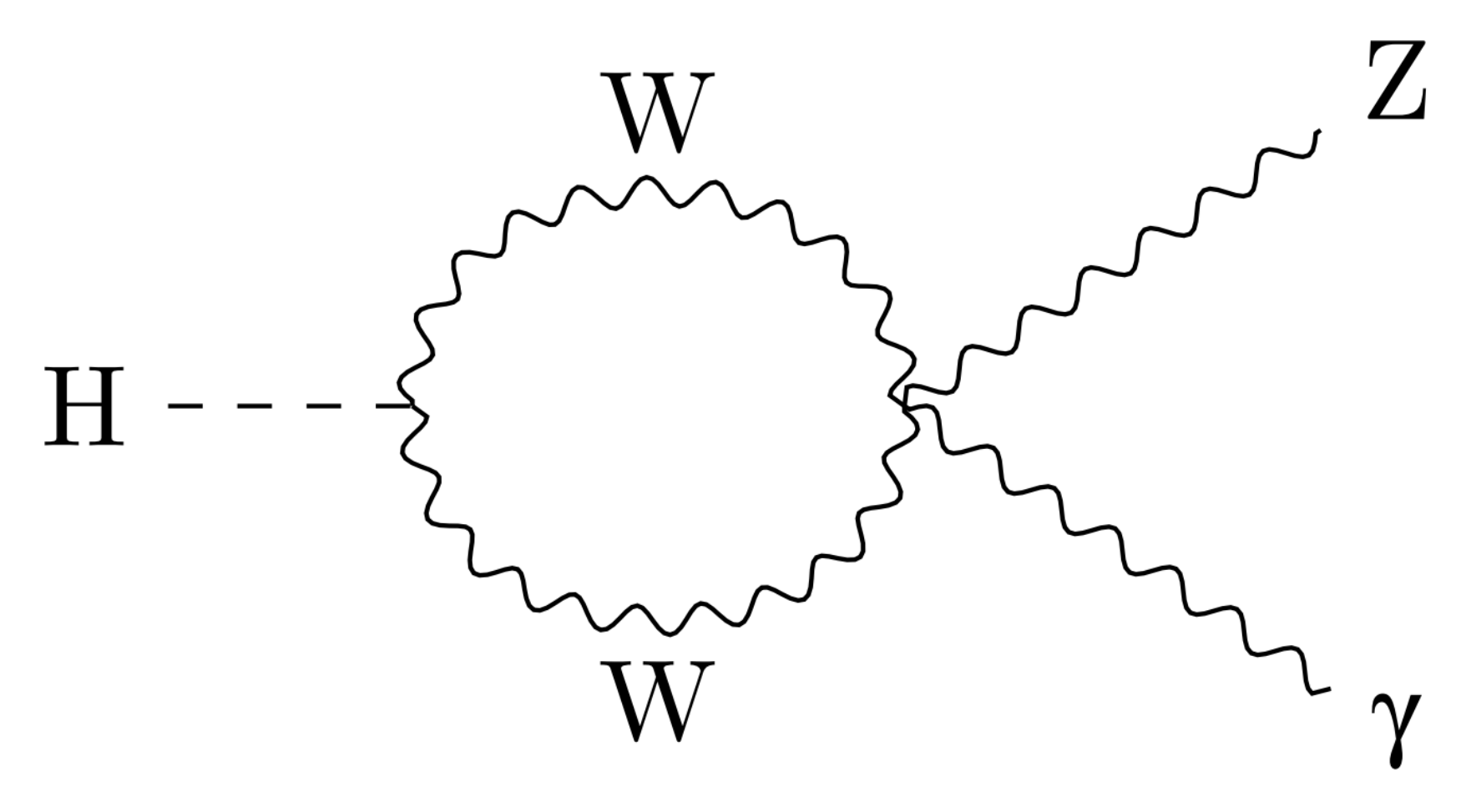}
  \includegraphics[width=0.3\textwidth]{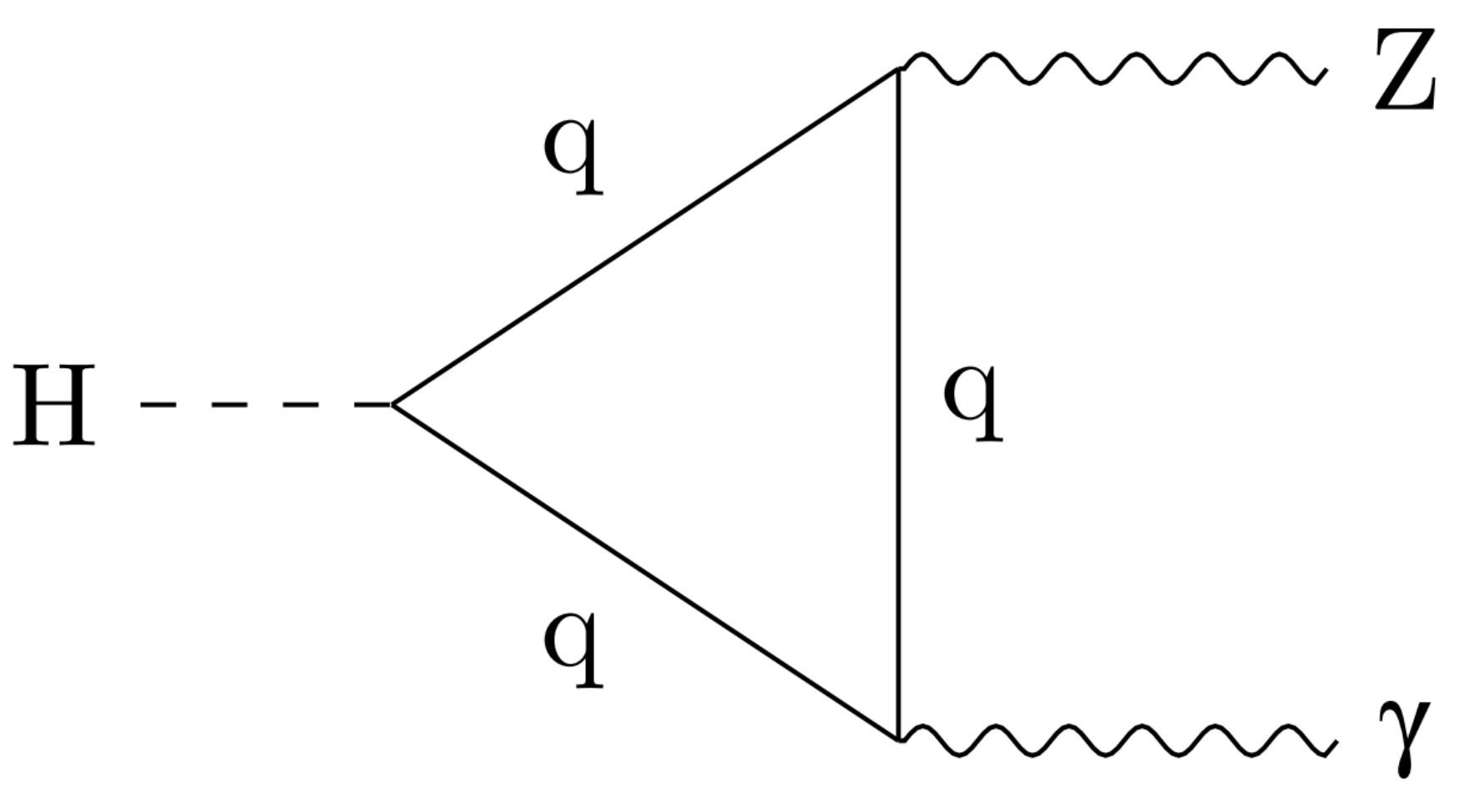}
  \caption{Feynman diagrams for $\HZG$ in the SM. In these loop-induced processes, the amplitudes associated with $\PW$ bosons in the
  intermediate state dominate over those with quarks.}
\label{fig:feynman}
\end{figure*}
}

The search presented here is based on the decay sequence \HZG with $\Zll$,
where $\Pell = \Pe$ or $\PGm$. The charged leptons provide efficient trigger paths, and the dilepton-plus-photon
final state yields a robust experimental signature. The pair of opposite-sign same-flavor leptons is required to have an invariant mass compatible with that of the $\PZ$ boson. 
The analysis separately considers Higgs boson production via gluon-gluon fusion (ggF), vector-boson fusion (VBF), and associated-production processes ($\ZH$, $\WH$, and $\ttH$), optimizing the sensitivity to each channel. Although ggF provides
the largest contribution to the total production
cross section, the other production channels offer additional distinctive experimental features and contribute to the signal sensitivity.
In all cases, the signal is extracted by fitting the distribution in the invariant mass of the final-state leptons and photon, $\mllg$, for a mass peak centered at the Higgs boson mass $\mH$.
Because the Higgs and $\PZ$ boson masses are similar, the photon energy in the Higgs boson rest frame is modest (approximately 30\GeV), and can be Lorentz boosted to even lower values in the laboratory frame.
As a consequence, the identification and measurement of photons with relatively low transverse momentum (\pt) for the CERN LHC environment, mainly in the range 15--35\GeV, play an important role in this measurement.

The measurement of the $\HZG$ signal in proton-proton (pp) collisions is complicated by large nonresonant backgrounds from other SM processes
that can produce the same reconstructed objects in the detector. In most channels, the dominant background
arises from $\Pp\Pp\to\PZ\PGg$, which is initiated both by gluon-gluon and quark-antiquark collisions.
Another substantial background arises from events containing a $\PZ$ boson and a reconstructed photon candidate that does not correspond to a genuine isolated photon originating from the primary parton-level interaction. 
These photon candidates can originate from 
(i) misreconstruction effects, including electrons without an associated track passing the electron veto algorithms; (ii) other processes from the primary $\Pp\Pp$ interaction including nonprompt decays, jet fragmentation, and multiparton interactions; or (iii) a different primary $\Pp\Pp$ interaction vertex (pileup). 
Processes (i)--(iii) are referred to collectively as $\zfake$ in this paper and become increasingly important at higher instantaneous luminosities, as discussed in Section~\ref{Sec-CMSDetector}.
Together, the $\PZ$ boson production processes with either a real or fake photon candidate constitute a substantial background in the $\mllg$ distributions.
This background can be reduced using a range of kinematic and object-identification variables but nevertheless remains large compared to the expected signal.
However, these backgrounds do not produce a peak in the $\mllg$ distribution at $\mH$. The mass resolution in $\mllg$ is therefore critical to achieving good sensitivity to the signal process.

Both the ATLAS~\cite{ATLAS:2014fxe,ATLAS:2020qcv} and CMS~\cite{CMS:2013rmy,CMS:2018myz,CMS:2022ahq} Collaborations at the LHC have reported results from
searches for $\HZG$ using this experimental signature. The ATLAS result using the 2015--2018 data set~\cite{ATLAS:2020qcv} and the CMS result using the 2016--2018 data set~\cite{CMS:2022ahq} each had a statistical significance below 3 standard deviations. 
A statistical combination of these two measurements yielded a signal strength $\mu$, defined as the ratio of the measured product of the cross section and decay branching fraction $\sigma(\ppH)\mathcal{B}(\HZG)$ to the corresponding SM prediction, of $\mu = 2.2\pm 0.6\stat^{+0.3}_{-0.2}\syst=2.2\pm0.7$~\cite{ATLAS:2023yqk}. 
The measured (expected) significance in the combination was found to be 3.4 (1.6) standard deviations. 
Assuming SM Higgs boson production cross sections, the resulting branching fraction  was $\mathcal{B}(\HZG) = (3.4 \pm 1.1)\times 10^{-3}$, corresponding to a difference of 1.9 standard deviations from the SM prediction.
However, a recent ATLAS update~\cite{ATLAS:2025aip}, including data collected in 2022--2024 in addition to their previous result, has reported a lower value of the signal strength, $\mu=1.3\ ^{+0.6}_{-0.5}\ $, with an observed (expected) significance of 2.5 (1.9) standard deviations. All existing measurements remain statistically limited.  

In this Letter, we present a new analysis based on CMS data samples
collected at $\sqrt{s} = 13\TeV$ in 2016--2018 (Run~2) and at
$\sqrt{s} = 13.6\TeV$ in 2022--2023 (Run~3).
The integrated luminosities for these two samples are 138\fbinv and
62\fbinv, respectively, for a total of 200\fbinv.
The increase in the collision energy from 13 to 13.6\TeV leads to an
increase in the SM Higgs boson production cross section of about
5\%~\cite{deFlorian:2016spz,Karlberg:2024zxx}.

This new analysis incorporates several improvements with respect to the previous CMS Run~2 measurement~\cite{CMS:2022ahq}.
The main enhancements include (i) a new trigger strategy that employs both single-lepton and dilepton triggers;
(ii) a new primary categorization scheme with additional categories targeting specific Higgs boson production processes; (iii) new object and event selection
criteria, increasing the signal efficiency in sensitive regions of phase space; 
(iv) redesigned and better targeted machine-learning algorithms for classification of events within the primary categories; (v) additional validation of the fit to the distribution in \mllg; and (vi) a more detailed treatment of the systematic
uncertainties in the \mllg fit. These developments are designed to increase
the expected signal significance and to improve the reliability of the results. To achieve optimal
sensitivity and to provide a uniform treatment of the data, these improvements are
applied to both the Run~2 and Run~3 data sets. 
The expected significance is improved by approximately 50\% relative to the expectation obtained by extrapolating the previous CMS measurement to the integrated luminosity of the data sample analyzed in this Letter.

The Letter is organized as follows. Sections~\ref{Sec-CMSDetector} and \ref{Sec-SimulatedEventSamples} describe the
CMS detector and the simulated event samples, respectively. The triggers, basic event reconstruction procedures, and
baseline event selections are described in Section~\ref{Sec-TrigAndEventReconstruction},
while the signal and background event properties, discriminating variables, and event categorization 
are presented in Section~\ref{Sec-EventSelection}.
Section~\ref{Sec-FittingAndResults} describes the methodology used for the signal extraction, including
the data models, tests, and validation procedures employed in the maximum likelihood fits.
Section~\ref{Sec-SystematicUncertainties} discusses the systematic uncertainties, and the
results are presented in Section~\ref{Sec-Results}. 
A summary is provided in Section~\ref{Sec-Summary}. 
Tabulated results are provided in the HEPData record for this analysis~\cite{hepdata}.

\section{The CMS detector\label{Sec-CMSDetector}}

The CMS apparatus~\cite{CMS:2008xjf,CMS:2023gfb} is a multipurpose, nearly hermetic detector, designed to trigger on~\cite{CMS:2020cmk,CMS:2016ngn,CMS:2024aqx} and identify electrons, muons, photons, and (charged and neutral) hadrons~\cite{CMS:2020uim,CMS:2018rym,CMS:2014pgm}. A global ``particle-flow'' (PF) algorithm~\cite{CMS:2017yfk} aims to reconstruct all individual particles in an event, combining information provided by the all-silicon tracker and by the crystal electromagnetic calorimeter (ECAL) and brass-scintillator hadron calorimeter (HCAL), operating inside a 3.8\unit{T} superconducting solenoid, with data from the gas-ionization muon detectors embedded in the flux-return yoke outside the solenoid. The reconstructed particles are used to build \PGt leptons, jets, and missing transverse momentum~\cite{CMS:2018jrd,CMS:2016lmd,CMS:2019ctu}. 

Events of interest are selected using a two-tiered trigger system. The first level, composed of custom hardware processors, uses information from the calorimeters and muon detectors to select events at a rate of around 100\unit{kHz} within a fixed latency of 4\mus~\cite{CMS:2020cmk}. The second level, known as the high-level trigger (HLT), consists of a farm of processors running a version of the full event-reconstruction software optimized for fast processing, and reduces the event rate to a few kHz before data storage~\cite{CMS:2016ngn,CMS:2024aqx}. 

In addition to the $\Pp\Pp$ collision of interest selected by the trigger, the event record
generally contains tracks and calorimeter energy deposits from a number of other
collisions (pileup) occurring in the same bunch crossing, or in adjacent crossings within
the time resolution of the data acquisition.  The distribution of the number of $\Pp\Pp$ interactions per bunch crossing varies with the
beam conditions and instantaneous luminosity, with a mean of about 29 over the Run~2 data~\cite{CMS:2020ebo} and increasing to 40 in 2022 and 45 in 2023, assuming an inelastic pp cross section of 69.2 mb.

\section{Simulated event samples\label{Sec-SimulatedEventSamples}}

Simulated event samples play several important roles in this analysis. 
First, they are used to optimize the event selection and event categorization criteria to maximize the
expected sensitivity to the signal processes. They are also used to design, test, and partially validate
the analysis methodology and the fitting method in particular.
Finally, they are used to determine the efficiency factors that relate the observed signal yield in each category to the number of produced signal events and the corresponding signal strengths.
Notably, the simulated event samples are not used in the background estimation itself, which is based on fits to the \mllg distributions in data. 
Simulated event samples for background processes are extensively validated to ensure their ability to describe all kinematic quantities used to discriminate between signal and background; these checks are performed using several control regions in the data. 

Simulated signal events are generated in the primary Higgs boson production channels: ggF, VBF, and the associated-production processes \VH (\WH or \ZH) and \ttH. 
For each production channel, samples are generated for Higgs boson mass values of $\mH=120$, 125, and 130\GeV to derive parametric signal models by interpolating the fit parameters obtained at these mass values.

The dominant nonresonant SM background contributions arise from \smzg and \zfake production, where in the latter process the fake photon can arise in several ways, as explained in Section~\ref{Sec-Introduction}. 
After the baseline selection described in the next section, these processes comprise about 60\% and 40\% of background events, respectively.
Additional SM background contributions arise from multiboson production, production of a \PZ boson via vector boson scattering (\vbszg), and processes involving top quarks (\tgttg, \tttw, and \ttX). 
Although these processes only comprise about 1\% of the total background integrated across all categories, they can be substantial in categories that target \VH or \ttH production.
A small resonant background arises from the process \Hmm in which one of the muons emits a photon via final-state radiation (FSR). 
This background amounts to about 3\% of the SM signal yield after the baseline selection.
This contribution, fixed to the expected SM yield, is explicitly included in the statistical model used in the fits. 

Simulation of the \HZG signal, \Hmm, \tttw, and some multiboson background processes is performed with the {\POWHEG}\,v2.0~\cite{Nason:2004rx,Frixione:2007vw,Alioli:2010xd,Re:2010bp,Frixione:2007nw,Alioli:2011as,Melia:2011tj,Nason:2013ydw,Kardos:2014dua,Alioli:2008tz,Bagnaschi:2011tu,Nason:2009ai,Luisoni:2013cuh,Hartanto:2015uka} generator at next-to-leading order (NLO) precision in perturbative quantum chromodynamics (QCD). Gluon-gluon fusion signal events are reweighted so that the generated Higgs boson \pt distribution matches an auxiliary sample generated at next-to-next-to-leading order (NNLO) precision using \MiNNLO interfaced with the \POWHEG event generator~\cite{Monni:2019whf,Monni:2020nks}.

Simulation of the \smzg, the \zfake, some multiboson, the \tgttg, and some \ttX samples is performed using the {\MGvATNLO}\,v2.6 (Run~2) or v2.9 (Run~3)~\cite{Alwall:2014hca,Alwall:2007fs} generator at NLO precision. The \textsc{MadSpin} module~\cite{Artoisenet:2012st} is used for the \tgttg samples, and the FxFx merging scheme~\cite{Frederix:2012ps} is employed when additional partons are included in the matrix element calculations.
The \vbszg and the remaining \ttX samples are generated with the \MGvATNLO\,v2.6 (Run~2) or v2.9 (Run~3) generator at leading order (LO) precision, with MLM jet matching employed in cases when extra partons are generated in the matrix element calculations~\cite{Alwall:2007fs,Mangano:2001xp,Mangano:2002ea}.
The remaining diboson samples are generated using \PYTHIA\,v8.240 (Run~2) or v8.306 (Run~3)~\cite{Sjostrand:2014zea} at LO precision.

All simulated samples use the NNPDF3.1 NNLO~\cite{Ball:2017nwa} parton distribution functions (PDFs). 
\PYTHIA is used to describe the parton shower and fragmentation, using the CP5 tune~\cite{CMS:2019csb} for the underlying event model.
The nominal CP5 tune is used for all samples except for a set of special signal samples with variations of the CP5 tune, which are used to assess the impact of the uncertainty in the underlying event model.
Detector simulation is performed with the \GEANTfour package~\cite{Agostinelli:2002hh}.
Finally, the pileup activity is incorporated by overlaying the simulated events with minimum bias events generated with \PYTHIA.

Simulated event samples are normalized to their cross sections multiplied by the integrated luminosity for each data-taking period.
The cross sections for Higgs boson production processes follow the recommendations of the LHC Higgs Working Group~\cite{deFlorian:2016spz,Karlberg:2024zxx}, computed at next-to-NNLO accuracy (ggF), NNLO accuracy (VBF and \VH), or NLO accuracy (\ttH) in QCD, together with NLO (ggF, VBF, and \VH) or LO (\ttH) accuracy in electroweak corrections.
The 13\TeV cross section for \PZ boson production (used for the \zfake sample) is derived at NNLO precision in QCD with \FEWZ 3.1~\cite{Li:2012wna}, and the cross sections for top quark pair production are obtained from NNLO QCD computations plus next-to-next-to-leading logarithmic soft-gluon resummation using the \textsc{Top++} program~\cite{Czakon:2011xx}.
Other cross sections are taken directly from the relevant LO or NLO event generators.
Additional event weights are applied to correct for observed differences between data and simulation.
For some nonresonant background samples, an overlap removal procedure is applied to avoid double counting events in the phase space simulated in multiple samples.

Because the cross section for \PZ boson production is large, it is computationally difficult to directly produce simulated event samples for \zfake events of sufficient size. 
We have therefore developed dedicated
techniques to increase the effective sample size. Events from the \zfake sample passing the baseline selection
(Section~\ref{Sec-TrigAndEventReconstruction}) are categorized into three classes based on the presence and types of generator-level particles within $\deltar=\sqrt{\smash[b]{(\Delta \eta)^2+(\Delta \varphi)^2}} < 0.1$ of the photon candidate. 
The first class, consisting of events with a generator-level photon in this cone, primarily contains photons produced in the parton shower radiation or from hadron decays; this class is supplemented using a sample with a generator-level photon requirement.
The second class, consisting of events without generator particles in this cone, originates from pileup photons and detector noise, and is supplemented with a sample generated by mixing lepton candidate pairs and photon candidates from different events in the existing samples.
The third class, consisting of events with generator particles other than a photon in the cone, is a mixture of several processes and is supplemented with a sample generated with a linear combination of the first two methods.
The extended \zfake samples are validated and found to reproduce the distributions of the original \zfake sample in all analysis variables.

\section{Triggers, event reconstruction, and baseline selection\label{Sec-TrigAndEventReconstruction}}

The data sample was collected using a logical OR of single-lepton and dilepton triggers, providing an increase in efficiency of approximately 10\% compared to dilepton triggers alone.
The dilepton triggers require the leading and subleading electrons (muons) to satisfy $\pt > 23~(17)\GeV$ and $\pt > 12~(8)\GeV$, respectively, as reconstructed by the HLT.
The single-electron and single-muon trigger $\pt$ thresholds vary depending on the data-taking conditions and  
are in the range 27--32\GeV (electrons) and 24--27\GeV (muons). 
The HLT also imposes loose lepton isolation criteria to help reduce the rate of events with leptons produced in
the decays of hadrons containing heavy quarks.
Trigger efficiencies are measured using the tag-and-probe method \cite{CMS:2011aa}, in which an unbiased sample of probe leptons from $\Zll$ decays
is obtained by requiring a separate tag lepton in each event to pass a single-lepton trigger. The trigger
efficiencies rise at $\pt$ values near the thresholds and reach an efficiency plateau in the range 95--99\%.
Because of the slightly higher lepton $\pt$ thresholds applied later, the analysis operates in
the plateau regions of the trigger efficiency versus lepton \pt turn-on curves.
Systematic uncertainties associated with the trigger selection efficiencies are discussed in Section~\ref{Sec-SystematicUncertainties}. 

The primary vertex (PV) is taken to be the vertex with the largest sum of associated physics-object $\pt^{2}$ values in the event, as described in Ref.~\cite{CMS-TDR-15-02}.
The tracks of charged particles associated with vertices other than the PV are removed from further consideration.
The PV is required to lie within 24\unit{cm} of the center of the detector in the direction along the beam axis
and within 2\unit{cm} in the plane transverse to that axis.

The signatures for signal events are based primarily on the reconstruction and measurement of electrons, muons, and photons from the Higgs boson decay chain.
However, depending on the Higgs boson production channel, additional objects, such as jets, $\Pb$-tagged jets, and FSR photon candidates, can be incorporated into the signature.
The reconstruction criteria used for all of these objects are discussed below.

Events collected using electron triggers must either pass the single-electron trigger and have an electron candidate with $\pt>30\GeV$ or $\pt>35\GeV$ depending on the trigger threshold, or pass the dielectron trigger and have leading and subleading electron candidates with $\pt>25\GeV$ and $\pt>15\GeV$, respectively.
All electron candidates are required to satisfy the selections on transverse momentum and pseudorapidity $\pt>7\GeV$ and $\abs{\eta}<2.5$. 
Electron candidate tracks must pass within 1.0\unit{cm} (0.5\unit{cm}) of the primary vertex in the beam axis (transverse) direction.
As the background from nonprompt leptons is very small in most analysis categories, only a loose requirement on a
multivariate identification/isolation discriminant is applied to all electrons.
In categories with additional leptons, a further requirement on a version of the mini-isolation variable \Imini~\cite{Rehermann:2010vq} of $\Imini<0.15$ or $\Imini<0.1$ is applied, as described in Section~\ref{Sec-EventSelection}.
The variable \Imini is defined as the sum of the transverse momenta of other PF candidates in a cone of $\deltar<R_0$, 
divided by the lepton transverse momentum \ptl, where the cone radius $R_0=0.2$ when 
$\ptl<50\GeV$, $R_0=10\GeV/\ptl$ when $50<\ptl<200\GeV$, and $R_0=0.05$ when $\ptl>200\GeV$.

Events collected using muon triggers must either pass the single-muon trigger and have a muon candidate with $\pt>25\GeV$ or $\pt>28\GeV$ depending on the trigger threshold, or pass the dimuon trigger and have leading and subleading muon candidates with $\pt>20\GeV$ and $\pt>10\GeV$, respectively.
All muon candidates are required to satisfy $\pt>5\GeV$ and $\abs{\eta}<2.4$.
Muon candidate tracks must pass within 1.0\unit{cm} (0.5\unit{cm}) of the primary vertex in the beam axis (transverse) direction, and must have a three-dimensional impact parameter significance less than four.
Muon candidates must satisfy loose cut-based identification requirements \cite{CMS:2018rym} and have relative isolation $\Irel<0.35$, where $\Irel$ is defined as the sum of the transverse momentum of other PF candidates within a cone of $\deltar<0.3$ divided by the transverse momentum of the muon. In categories with additional leptons, a selection on mini-isolation is applied, as for electrons. 

Photon candidates are required to satisfy $\pt>15\GeV$, $\abs{\eta}<2.5$, and to fall outside the ECAL barrel-endcap transition region.
Identification is performed by applying a selection on a multivariate discriminator with a nominal 80\% efficiency for genuine photons in a \gjets simulated sample, and a veto is applied to suppress electrons that do not originate from a photon conversion~\cite{CMS:2020uim}.
Photon candidates with $\deltar(\Pell,\PGg)<0.3$ are rejected, where $\Pell$ refers to any electrons or muons passing their respective selections.
Finally, photon candidates that also satisfy the electron selection criteria are removed. 

While small-angle FSR photons are generally included in the calorimeter clustering procedure for electron reconstruction, they are not considered in muon reconstruction.
We therefore apply an FSR recovery procedure for muons using
FSR photon candidates satisfying $\pt>2\GeV$ and $\abs{\eta}<2.4$.
The FSR photons must satisfy $\deltar(\mu,\PGg)/\ptg^{2}<0.012\GeV^{-2}$, and have a relative isolation $\Irel<1.8$.
Finally, FSR photons must satisfy $\deltar>0.2$ with respect to the photon
selected for reconstructing the candidate Higgs boson system.
Approximately 2\% of muon candidates are associated with FSR photons. 
The FSR photons associated with muon candidates are then incorporated in a fit procedure described in Section~\ref{Sec-EventSelection} to improve the precision of muon four-momentum reconstruction. 

Although jets are not needed to reconstruct the Higgs boson decay, they can be important parts of the signature, particularly in the VBF and \ttH channels, which can be distinguished by the presence of forward and \PQb-tagged jets, respectively.
For this purpose, jets are reconstructed from PF candidates using the anti-\kt algorithm~\cite{Cacciari:2008gp} with radius $R=0.4$ as implemented in the \FASTJET package~\cite{Cacciari:2011ma}, and pileup mitigation corrections are applied using the charged-hadron subtraction technique~\cite{Cacciari:2007fd} for Run~2 data or the pileup-per-particle identification (PUPPI)~\cite{Bertolini:2014bba, CMS:2020ebo} algorithm for Run~3 data.
Spurious jets are suppressed using a set of jet identification criteria~\cite{CMS:2020ebo}, and the jet candidates are required to satisfy $\pt>30\GeV$ and $\abs{\eta}<4.7$.
Because of mismodeling in simulation, jets with $\pt<50\GeV$ and $2.5<\abs{\eta}<3.0$ are vetoed. 
The identification of jets arising from the hadronization of a $\PQb$ quark and its subsequent fragmentation is important
for studying \ttH production. 
The \textsc{DeepJet} neural network $\PQb$ tagger~\cite{Bols:2020bkb} is used to identify such jets. 
Jets satisfying the DeepJet medium working point criteria are considered to be b-tagged jets.
In simulated top quark events, approximately 80\% of genuine $\PQb$ jets are correctly identified, with a 1\% mistag rate for non-$\PQb$ jets~\cite{CMS-DP-2023-005,CMS-DP-2023-012,CMS:2017wtu}.
The missing transverse momentum (\ptmiss) is computed as the negative vector sum of the \pt of all PF candidates in an event~\cite{CMS:2019ctu}. In the Run~3 data sets, this sum is weighted using the PUPPI algorithm~\cite{Bertolini:2014bba,CMS:2020ebo}.
Jet energy corrections applied to simulation are propagated to \ptmiss~\cite{CMS:2016lmd,CMS-DP-2021-033}. 

A baseline event selection is applied to all samples, both in data and simulation, prior to event classification, categorization procedures, validation, and fitting.
This selection consists of the requirements on the reconstructed objects described above, together with the following set of loose kinematic criteria: 
\begin{enumerate}
\item A single-lepton or dilepton trigger must be satisfied, together with the associated offline lepton \pt selections.
\item The event must contain at least two opposite-sign same-flavor (OSSF) leptons satisfying the criteria described above. If an event contains more than one OSSF lepton pair, the candidate system with invariant mass closest to that of the $\PZ$ boson is chosen.
\item The event must contain at least one isolated photon satisfying the criteria described above. The photon with the highest \pt is then used to form the $\llg$ system; fewer than 0.1\% of simulated signal events contain more than one photon candidate.
\item The invariant mass $\mll$ of the OSSF dilepton system must satisfy the requirement $80 < \mll < 100\GeV$. This rejects almost all events that do not contain the processes \Zee or \Zmm.
\item The candidate systems must satisfy the selection on the ratio $\ptg/\mllg > 15/110 $ to suppress the $\zfake$ background while maintaining a smooth falling distribution of background in \mllg near \mH.
\item The candidate systems must satisfy $\mllg+\mll > 185\GeV$ to suppress background in the region $\mllg < 105\GeV$ from FSR \Zllg production, which peaks strongly at $\mllg=91\GeV$ and can make the distribution of background in \mllg more difficult to model.
\item Events with spurious sources of $\ptmiss$ are removed using a set of event filters~\cite{CMS:2019ctu}.
\end{enumerate}  

The baseline selection has an average efficiency of about 27\% for signal events, with small variations across the production modes.
After the baseline selection, about 280 signal events are expected in the ggF channel, 23 in the VBF channel, and 14 in the associated-production channels combined. 
However, the background yield in the Higgs boson mass region ($120<\mllg<130\GeV$) is still of order $10^5$ events overall. Therefore, multivariate event classifiers are used to further improve the analysis sensitivity.

\section{Signal and background properties; event categorization\label{Sec-EventSelection}}

After the baseline selection (Section~\ref{Sec-TrigAndEventReconstruction}), the analysis proceeds in four main steps:
\begin{enumerate}
\item A \textit{primary event categorization} is applied, assigning events to categories according to object multiplicity requirements and kinematic features indicative of the different Higgs boson production channels. 
\item A \textit{secondary event categorization}, performed using boosted decision trees (BDTs),
  is applied in the ggF and VBF categories to improve the sensitivity to the signal. 
\item A \textit{$\PZ$-line-shape-constrained fit} is performed to improve the $\PH\to\llg$ mass resolution in the signal extraction.
  This fit uses the reconstructed four-momenta of the leptons making up the candidate $\Pellp\Pellm$ system and improves their values by constraining the dilepton invariant mass within the uncertainty of the measurements to a function describing the $\PZ$ boson natural line shape.  
\item \textit{Extraction of the signal event yields} is performed using a maximum likelihood fit to the \mllg distributions in the various event categories.
\end{enumerate}  
This section describes the first three of these steps, while the last step is discussed in Section 6.

Every event is assigned to exactly one of the primary event categories defined in Table~\ref{tab:PrimCatDefs}. These
categories are designed to target a specific Higgs boson production channel
and to improve the expected sensitivity by using quantities relevant to that channel.
The primary event categories are defined using the basic reconstructed physics objects including 
(i) the number of leptons in the event (\nl),
(ii) the number of jets in the event (\njet), (iii) the number of $\PQb$-tagged jets (\nbm),
(iv) the magnitude of the missing transverse momentum (\ptmiss), and (v) a small number
of category-specific quantities. These last quantities are listed under \textit{Binning/optimization selections} 
in Table~\ref{tab:PrimCatDefs} and are $\mll$, $\ptunderllg/\mllg$, and the \Imini selection applied to all leptons.
The primary categories combine events from all data-taking periods.

\begin{table*}[ht!]
  \centering
  \caption{Primary event categories defined using selections on leptons, jets, \ptmiss, and auxiliary variables. Each event is assigned to one of these nonoverlapping categories after the baseline selection is applied. In the case of the ggF and VBF categories, a further secondary categorization is subsequently applied using a boosted decision tree classifier (BDT). An event is assigned to the \tthlep category if it passes one of the two object criteria listed in the table.
  }
  \begin{tabular}{ l  c  c  c  c  c  c  c  c }
      \multirow{2}{*}{Category} & \multicolumn{4}{c}{ Lepton, jet, and \ptmiss requirements} & \multicolumn{3}{c}{Binning/optimization selections} \\
      \cline{2-8}
                  & \nl & \njet & \nbm  & \ptmiss (GeV) &  \mll (GeV) & $\ptunderllg/\mllg$    &  \Imini \\
      \hline
      ggF         & $=$2       & $\leq$1    &                      & $<$90         & \multicolumn{3}{c}{[Binned by ggF BDT (4 bins)]} \\
      [\cmsTabSkip]
      VBF         & $=$2       & $\geq$2    & $=$0                 &               & \multicolumn{3}{c}{[Binned by VBF BDT (4 bins)]} \\
      [\cmsTabSkip]
      \vhthreelep & $\geq$3   &            & $=$0                 & $>$30         &                       & $>$0.3            & $<$0.15  \\
      [\cmsTabSkip]
      \vhptmiss   & $=$2       & $\leq$1    &                      & $>$90        &                       & $>$0.4            & \\
      [\cmsTabSkip]
      \multirow{2}{*}{\tthlep} & $=$3      & $\geq$3    & $\geq$1              &               &                       & \multirow{2}{*}{} & \multirow{2}{*}{$<$0.1}\\
                  & $\geq$4   & $\geq$1    & $\geq$1              &               &                       &                   & \\
      [\cmsTabSkip]
      \tthhad     & $=$2       & $\geq$5    & $\geq$1              &               & (85, 95)              &                   & \\
      [\cmsTabSkip]
      Untagged & \multicolumn{7}{l}{Events passing baseline selection not assigned to any other category} \\
  \end{tabular}
  \label{tab:PrimCatDefs}
\end{table*}

Most of the signal events are expected to be in the ggF- or VBF-targeted primary categories. 
As shown in Table~\ref{tab:PrimCatDefs}, the ggF category allows at most one jet, while the VBF category requires at least two non-\PQb-tagged jets. 
The invariant mass of the dijet system, $\mjj$, is used as part of the multivariate classifier for the VBF category, as discussed below.
To enter the categories targeting the \ttH production mode, events must contain at least one $\PQb$-tagged jet in addition to a requirement on the total number of jets that depends on the top quark decay mode.
In the categories targeting the $\ZH$ and $\WH$ production modes, an additional
lepton or a minimum value of the $\ptmiss$ variable is required, reflecting the presence of charged leptons and/or neutrinos
in leptonic $\PW$ and $\PZ$ boson decays.
Events not assigned to a category described above are assigned to the untagged category.

After the baseline selection, approximately 80\% of all signal events in simulation are assigned to the ggF category, approximately 15\% to the VBF category, and approximately 5\% to other categories.
While VBF events comprise only about 25\% of the events in the inclusive VBF category, this fraction is enhanced in some of the secondary categories, as described below.

A secondary categorization is then applied to events in the ggF and VBF primary categories. 
Two BDTs, one for ggF and one for VBF, are trained with the \textsc{XGBoost} package~\cite{Chen:2016:XST:2939672.2939785} using simulated signal and background
event samples (Section~\ref{Sec-SimulatedEventSamples}). 
The BDTs are trained with a set of variables that characterize the quality of the reconstructed objects, their measurement uncertainties, and the event kinematic properties, as discussed below. 
The variables are chosen based on their ability to provide a statistically significant enhancement to the estimated sensitivity. 
Furthermore, the variable selection is tuned to avoid sculpting the distribution of background in \mllg to peak near \mH, though small correlations between BDT input variables and \mllg are permitted.
Each BDT then yields a single discriminating variable, or score, that is used for binning the events in that primary category.
In both the ggF and VBF primary categories, simulated background events for the $\smzg$, $\zfake$, and $\vbszg$ processes are included in the training procedure. 
For the purposes of training the BDTs, all events are restricted to lie within the Higgs boson mass region.

The difficulty of preparing large simulated event samples for the backgrounds is addressed by using a four-fold cross-validation procedure and a set of specially developed algorithms to extend the $\zfake$ sample as described in Section~\ref{Sec-SimulatedEventSamples}. 
Although the \zfake background is suppressed by the quality requirements used in the baseline selection, this background remains substantial, so a large event sample is important for training.
In the four-fold cross-validation procedure, the simulated samples are divided into four subsets (folds), and the BDT training procedure is repeated, each time using two folds for training, one for validation, and one for testing, with the roles rotated through the four folds, producing four separate BDTs trained and applied to different folds.
The training subsample is used for model training; the validation subsample is used for hyperparameter optimization, bin optimization, and other validation; and the testing subsample is used for final evaluation. 
These procedures ensure that, as the same BDT is never evaluated on the events used for its own training or optimization, the signal efficiency used to extract the signal strength is not biased.

After training, events are binned in BDT score. The bin boundaries are selected by maximizing the sum in quadrature of the estimated cut-and-count significance \begin{equation}
  \expsig = \sqrt{{2[(\nsig + \nbkg)\ln(1+\nsig/\nbkg) - \nsig]}},
\end{equation}
where $\nsig$ and $\nbkg$ refer to the numbers of simulated signal and background events from the validation subsample within the Higgs boson mass region.
A minimum number of simulated events is required in each bin to ensure that the fit to the distribution in $\mllg$ remains stable.
The categories are numbered ggF1--ggF4 and VBF1--VBF4, where bin 1 has the highest signal-to-background ratio and bin 4 has the lowest. 
These 8 secondary categories, together with the 5 remaining primary categories, make up the 13 categories that enter the statistical analysis.

The variables used in training the BDTs for the ggF and VBF categories are listed in Table~\ref{tab:ggfbdt-inputs}, where they are grouped
into the main types according to Higgs boson kinematic variables, photon quality indicators, angular variables,
jet variables, and system-balance indicators.
Here we highlight some of the key quantities.
The photon identification variable, IDMVA, incorporates information about the photon shower shape and isolation that characterize the quality of the candidate photon~\cite{CMS:2020uim}; it is used in conjunction with the photon energy resolution, $\sigmarel$.
As noted above, all photons must satisfy a minimal quality criteria as part of the baseline selection. 
However, even after that selection, the IDMVA variable is still useful in distinguishing signal from background events and is included in both the ggF- and VBF-targeted BDTs.

\begin{table*}[tb!]
\centering
\topcaption{Input variables for the BDT used in the ggF and VBF event categories. More details on these variables are given in Section~\ref{Sec-EventSelection}.}
\label{tab:ggfbdt-inputs}
\cmsTableIf{
\begin{tabular}{c  l l l}
\hline
Input variable & Description & ggF & VBF\\
\hline
	\multicolumn{4}{c}{Higgs boson kinematic variables} \\
\hline
	$\ptunderllg/\mllg$ &  Higgs boson candidate $\pt$ relative to its mass & \checkmark & \checkmark \\
\hline
	\multicolumn{4}{c}{Photon quality indicators}\\
\hline
	\PGg IDMVA & Photon identification MVA score  & \checkmark & \checkmark \\
	\sigmarel & Photon energy resolution  & \checkmark & \checkmark\\
\hline
\multicolumn{4}{c}{Angular variables of final state particles from Higgs boson candidate}\\
\hline
	\etalo, \etalt, \etag & Pseudorapidity of leptons and photon  & \checkmark & \checkmark \\
	\drmin and \drmax & Separation between photon and leptons  & \checkmark & \checkmark\\
	$\cos \Theta$ & \PZ boson production angle  & \checkmark & \checkmark\\
	$\cos \theta$ & \PZ boson decay polar angle  & \checkmark & \checkmark\\
	$\phi$ & \PZ boson decay azimuthal angle  & \checkmark & \checkmark\\
\hline
	\multicolumn{4}{c}{Jet variables}\\
        \hline
        $\eta(\text{j})$ & Jet pseudorapidity  & \checkmark &     \\
        $\pt(\text{j}_{1,2})$ & Jet transverse momenta  & \checkmark & \checkmark\\
        \detajj & Dijet pseudorapidity separation  &     & \checkmark\\
        \dphijj & Dijet azimuthal angular separation  &     & \checkmark \\
        $\deltar(\text{j}_{1,2},\PGg)$ & Jet separation from photon  & \checkmark & \checkmark\\
        $m_\text{j}$, \mjj & Jet/dijet system invariant mass  & \checkmark & \checkmark\\
        \njet & Number of jets  & \checkmark & \checkmark\\
\hline
        \multicolumn{4}{c}{System balance indicators}\\
\hline
        \dphillgj, \dphillgjj & Azimuthal separation between jet/dijet and $\llg$  & \checkmark & \checkmark\\
        \sysbalance & System \pt balance  & \checkmark & \checkmark\\
$ \abs{\etag-\eta(\text{j})}$, \zeppenfeld & Pseudorapidity separation between photon and jet(s) & \checkmark & \checkmark\\
        \dphiphstmiss & Azimuthal angular separation between photon and \stmiss  & \checkmark & \checkmark\\
\hline
\end{tabular}}
\end{table*}

The kinematic properties of Higgs boson production and the subsequent
two-body decay processes are well understood~\cite{Gainer:2011aa} and modeled in simulation, while those of the backgrounds are extensively studied in
both simulation and control regions in data. 
The backgrounds include a large contribution from $t$-channel processes, which lead to a more forward production of the particles in the $\llg$ system than is typical in signal events. 
The production angle of the $\PZ$ boson candidate with respect to the colliding partons in the Higgs boson rest frame, $\cos\Theta$, is therefore a useful discriminating variable.
Because it is a spin-0 particle, the Higgs boson must decay isotropically in its rest frame; furthermore,
the spin projection of the $\PZ$ boson along the $\HZG$ decay axis in the Higgs boson rest frame
must cancel that of the photon, forcing the $\PZ$ boson to have helicity $\pm 1$. 
This effect is reflected in the distribution of $\cos\theta$, the polar angle between the negatively charged lepton and the photon in the $\PZ$ boson rest frame.
While none of these kinematic quantities individually can provide powerful discrimination against the dominant backgrounds, the BDT is able to use them together to substantially increase the sensitivity of the analysis.

Although there is no requirement for jets to be present in the ggF category, the number of jets and the jet kinematic properties
for events with one jet are also provided to the BDT, as are 
variables that characterize the momentum balance of the system. 
Discrimination against reconstructed photons not associated with the PV is provided by the azimuthal angle between the photon candidate and $\stmiss$, defined as the negative vector sum of the $\pt$ of the reconstructed leptons, photons, and jets.
The most powerful variables in the BDT targeting the ggF category are the photon IDMVA variable, the photon relative energy resolution, $\drmin$, $\cos\Theta$, and $\ptunderllg/\mllg$.

The VBF process is characterized by the production of two jets with a large pseudorapidity gap, and quantities related to the individual jets and to the dijet system are included
in addition to the kinematic quantities used for the ggF-targeted BDT. The most important variables in the VBF-targeted BDT are  $\ptunderllg/\mllg$, $\drmin$, $\drmax$, $\mjj$, and \dphillgjj, where jj represents the dijet system. 

The modeling of kinematic and object quality variables is validated using data-simulation comparisons in four control regions: (i) a \Zllg region, (ii) a low quality photon region, (iii) a low \pt photon region, and (iv) the Higgs boson mass sideband regions (90--120, 130--180\GeV).
The distributions of the individual variables used in the BDTs are in good agreement between data and simulated background event samples in the Higgs boson mass sideband regions.
Following BDT training and optimization, validation studies are performed to confirm that the training is consistent across folds, the ranking of the variables in terms of discrimination is consistent with expectations, and there is minimal overtraining. 

\begin{figure*}[htb!]
 \centering
    \includegraphics[width=0.49\textwidth]{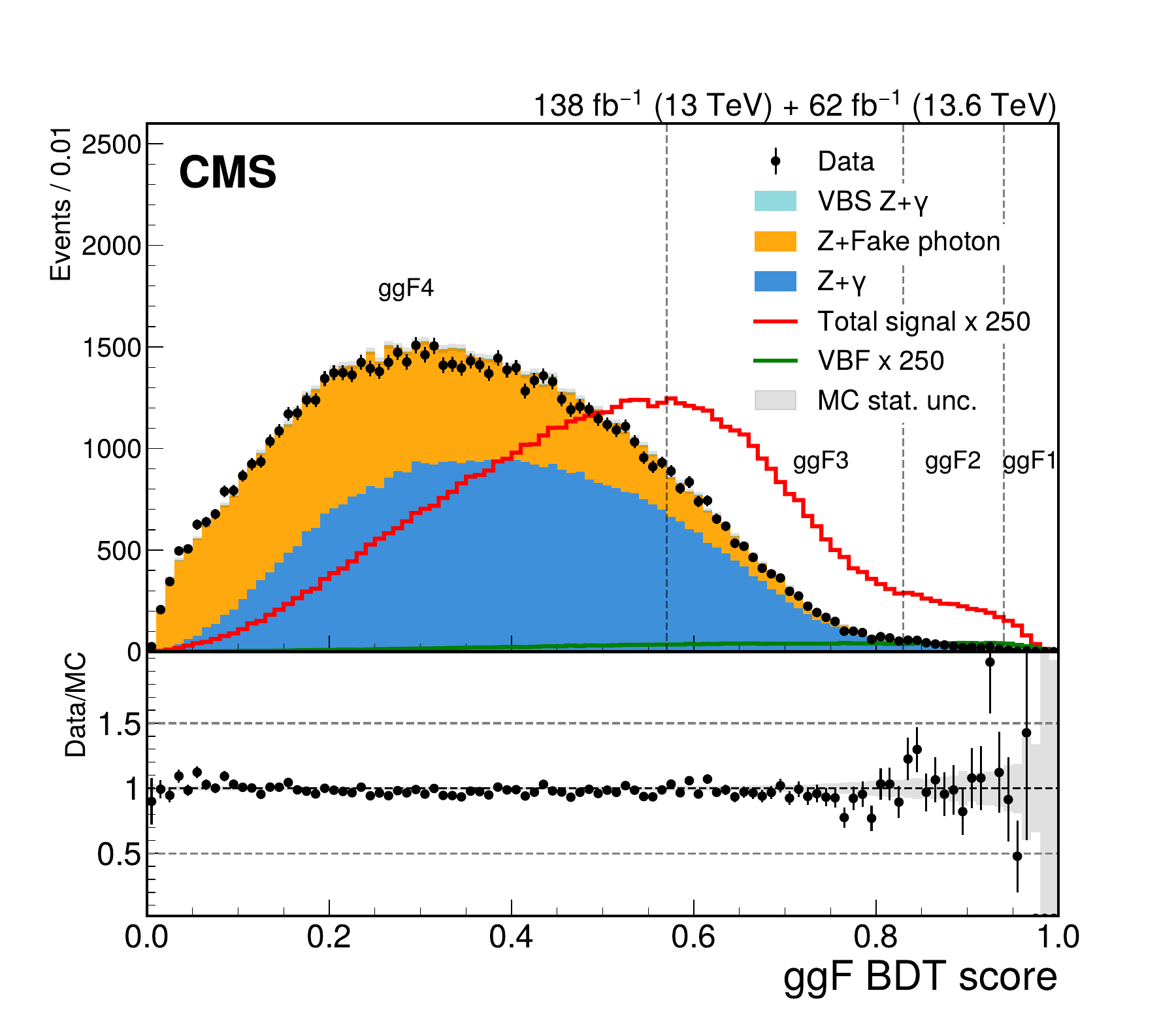}
    \includegraphics[width=0.49\textwidth]{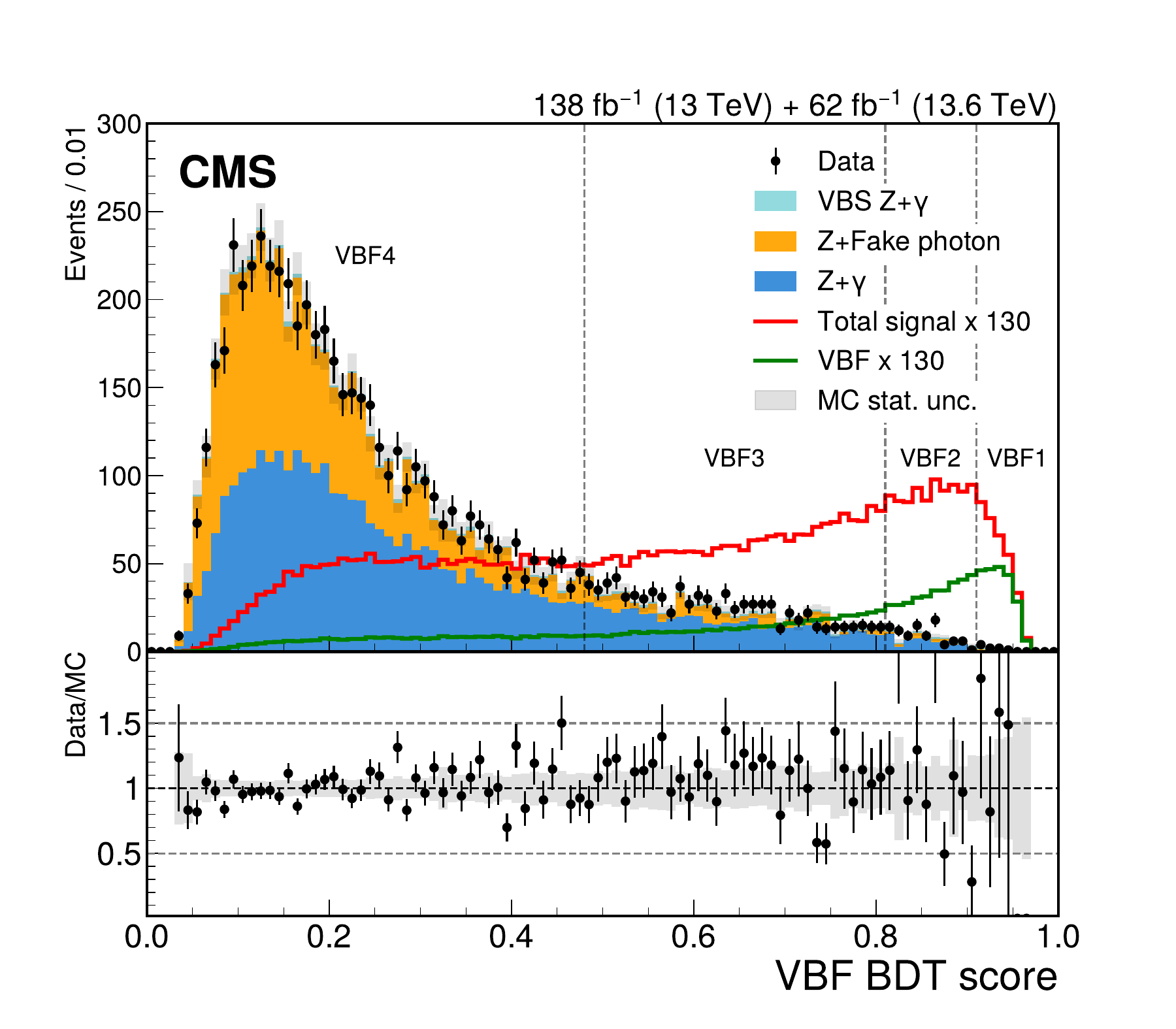}
  \caption{
    Distributions in BDT score for the (left) ggF and (right) VBF primary event categories restricted to the Higgs boson mass region (120--130\GeV).
    The data are shown as black points, the total simulated signal samples are shown as a red line, and the simulated background samples used in BDT training and evaluation are shown as filled histograms. 
    The subset of signal events from VBF production is also displayed independently as a green line.
    The sum of background MC samples are normalized to match data yields in the Higgs boson mass sideband regions.
    The signal event yields are scaled by a factor of 250 on the left plot, and 130 on the right plot for visibility. 
    The optimized BDT score bin boundaries are shown as vertical dashed lines. 
  }
    \label{fig:bdt_score}
\end{figure*}

Figure~\ref{fig:bdt_score} shows the distributions of data and simulated events in BDT score for the ggF and VBF primary categories.
The plots show that the distributions in the BDT scores are in good agreement between data and simulated background in the Higgs boson mass region.

The final step before performing the fits to obtain the signal yields using the $\mllg$ distributions is the $\PZ$-line-shape-constrained fit.  
Because the nonresonant SM backgrounds in the $\llg$ final state are large compared to the expected signal yields, the resolution in the reconstructed $\PH\to\llg$ invariant mass plays an important role in achieving the best sensitivity to the signal processes.
To improve the mass resolution, a maximum likelihood fit is performed that constrains the dilepton invariant mass using the known $\PZ$ boson line shape, as modeled in simulation with a total decay width of $\Gamma = 2.5\GeV$. 
The inputs to the fit are the lepton four-momenta, expressed in terms of the lepton $\pt$ values and angular quantities, as well as the associated uncertainties. 
The $\pt$ uncertainties are large compared to the angular uncertainties, which are ignored, so the fit varies only the $\pt$ values. 
For muon events, FSR photons and their associated $\pt$ uncertainties are also included in the fit.
The outputs of the fit are the improved estimates of the lepton $\pt$ values, which are used to recalculate the lepton four-momenta and the candidate Higgs boson mass used in the subsequent stages of the analysis.
The constrained fit provides a 10--20\% improvement in resolution, with a larger effect for $\Zee$ events than for $\Zmm$ events.

Table~\ref{tab:yields} summarizes the expected and observed event yields, effective signal widths, and estimated significance values in the primary and BDT-based secondary categories. 
The event yields are based on the event content in the Higgs boson mass region.
The table provides the estimated significance values in terms of \expsig and the effective signal width in terms of \sigmaeff, defined as half the width of the smallest interval containing 68.3\% of signal events, after applying the $\PZ$-line-shape-constrained fit. 
Higher purity BDT categories have better photon resolution and thus better \sigmaeff, as both the \sigmarel as well as the highly correlated photon IDMVA variable are used by the BDT to discriminate against the \zfake background.
The expected event yields are split into background and signal contributions, with the signal further split by production mode.
The signal in the four ggF categories is dominated by the ggF production mechanism. 
While the VBF categories are enhanced in VBF signal events relative to the ggF categories, only the highest purity bin of the VBF category contains signal events that are predominantly from VBF production (67\%) rather than from ggF.
In the associated-production categories, the fraction of signal events from the targeted production category is in the range 63--95\%.
The expected significance values show the importance of the VBF categories to the analysis sensitivity. 

\begin{table*}[tbp]
    \centering
	\caption{Summary of event categories with expected event yields, effective signal widths \sigmaeff, and estimated expected significance values \expsig based on simulated event yields. All yields are evaluated in the Higgs boson mass region 120--130\GeV. 
	The expected significance values are added in quadrature to obtain combined significance values that approximate the use of a statistical model combining the various categories.
	Because this method does not use information about the distribution in \mllg, the values obtained are lower than the expected significance using the fitting procedure described in Section~\ref{Sec-FittingAndResults}.}
    \cmsTable{
    \begin{tabular}{m{10em}ccccccccccccc}
    \hline
    Category &  ggF1 & ggF2 & ggF3 & ggF4 &
                         VBF1 & VBF2 & VBF3 & VBF4 &
                         \vhthreelep & \vhptmiss &
                         \ttH had. & \ttH lep. & Untag.\\
    \hline
    Expected signal yield & & & & & & & & & & & & & \\
    ~~~~ggF & 1.2 & 8.3 & 78 & 139 & 0.8 & 4.4 & 12 & 12 & 0.1 & 0.2 & 0.1 & 0.0 & 3.6\\
    ~~~~VBF & 0.5 & 1.8 & 4.0 & 3.5 & 1.7 & 2.6 & 3.6 & 2.3 & 0.0 & 0.1 & 0.0 & 0.0 & 0.8\\
    ~~~~\VH  & 0.0 & 0.3 & 1.5 & 1.9 & 0.0 & 0.4 & 1.5 & 1.8 & 1.0 & 0.5 & 0.0 & 0.0 & 1.5\\
    ~~~~\ttH & 0.0 & 0.0 & 0.0 & 0.0 & 0.0 & 0.0 & 0.1 & 0.1 & 0.1 & 0.0 & 0.7 & 0.6 & 1.0\\
    ~~~~Total & 1.7 & 10 & 84 & 144 & 2.5 & 7.5 & 17 & 16 & 1.2 & 0.8 & 0.8 & 0.6 & 6.9\\
    \\
    Expected bkg. yield & 16 & 311 & 9888 & 65493 & 5 & 82 & 721 & 4856 & 59 & 77 & 30 & 6 & 1683\\
    Data yield & 12 & 350 & 9787 & 64432 & 9 & 77 & 739 & 5058 & 77 & 92 & 33 & 7 & 1654\\
    \\
    $\sigmaeff(\eeg)$ (\GeVns) & 1.47 & 1.59 & 1.65 & 1.83 & 1.45 & 1.53 & 1.63 & 1.80 & 1.70 & 1.70 & 1.56 & 1.62 & 1.69\\
    $\sigmaeff(\mmg)$ (\GeVns)               & 1.23 & 1.26 & 1.29 & 1.50 & 1.29 & 1.32 & 1.36 & 1.50 & 1.47 & 1.34 & 1.34 & 1.38 & 1.42 \\
    \\
    \expsig & 0.4 & 0.6 & 0.8 & 0.6 & 1.0 & 0.8 & 0.6 & 0.2 & 0.2 & 0.1 & 0.1 & 0.2 & 0.2\\[\cmsTabSkip]
    Combined \expsig & \multicolumn{4}{c|}{1.2} & \multicolumn{4}{c|}{1.5} & \multicolumn{5}{c}{0.4} \\[\cmsTabSkip]
    Total \expsig & \multicolumn{13}{c}{2.0} \\
    
    \hline
    \end{tabular}}
    \label{tab:yields}
\end{table*}

\section{Fitting procedure, background estimation, and signal extraction\label{Sec-FittingAndResults}}

In the last major step of the analysis, we perform a simultaneous parametric maximum likelihood fit to the binned distributions in $\mllg$ across the 13 event categories using the CMS statistical analysis tool \textsc{Combine}~\cite{CMS:2024onh}.
In each category, the fit model includes three contributions: (i) the $\HZG$ signal; (ii) the combined nonresonant background, described by a single parametric function per category; and (iii) a small resonant contribution from $\Hmmgfsr$. 

The signal contribution in the $\mllg$ distribution is described separately in each category by a model based on a double-sided Crystal Ball (DSCB) function with a symmetric Gaussian core~\cite{Oreglia:1980cs,Gaiser:1982yw}, whose parameters and uncertainties are obtained from a fit to events in simulated samples for the signal processes (Section~\ref{Sec-SimulatedEventSamples}), combined across data-taking conditions.
The signal samples at 120, 125, and 130\GeV are used to interpolate the signal model to 125.38\GeV, which is taken as the nominal Higgs boson mass~\cite{CMS:2020xrn}.
The procedures used for the signal model are also applied in modeling the resonant background from $\Hmmgfsr$, which is constrained to its predicted yield in the SM~\cite{deFlorian:2016spz} within uncertainties. 
The yield from $\Hmmgfsr$ is about 2.7\% of that predicted for the \HZG signal.

The nonresonant backgrounds are modeled using empirical functions. 
To estimate the uncertainty in the background shape, the discrete profiling method~\cite{Dauncey:2014xga} is employed, whereby a set of candidate functions is prepared for each category, and the fit is allowed to select a preferred function via a discrete nuisance parameter.  Thus, the selected best fit function can vary with the signal strength $\mu$, propagating the uncertainty in the background shape to an uncertainty in $\mu$.
The selection of the set of background functions is described below.

In the ggF and VBF categories, the nonresonant background in the $\mllg$ distribution arises from two main contributions, as discussed in Section~\ref{Sec-EventSelection}: nonresonant $\smzg$ production and $\zfake$ production.
The distributions associated with both of these contributions involve a turn-on behavior at a threshold value of $\mllg$, a rapid increase to a peak region,
and a slow falloff at large values of $\mllg$. 
The generic turn-on behavior starts at a value around $\mllg = 95\GeV$ and is a consequence of the kinematic requirements.
The fit range is optimized independently for each category to include the peak region needed to constrain the change in the sign of the second derivative below the Higgs boson mass region as well as the tail that aids in constraining the slow falloff.
Two classes of parametric models to describe the sum of these nonresonant background contributions are considered: (i) convolution models, which are based on a function consisting of a sum-of-Gaussian core probability density function (pdf) convolved with the product of a step function and a set of functions to describe the falloff, and (ii) Gaussian-variant models, which employ direct modifications to the Gaussian function itself and are discussed below. 

The general mathematical expression for the convolution models is
\ifthenelse{\boolean{cms@external}}
{ 
\begin{multline*}
  \mathcal{F}(\mllg; \vec{\mu}_G, \vec{\sigma}_G, s, \vec{\alpha})\\ \equiv \int_{-\infty}^{\infty} \rd{t} \bigg(\mathcal{N}(\mllg - t; \vec{\mu}_G, \vec{\sigma}_G)
  \Theta(t;s)f(t;\vec{\alpha})\bigg),
    \label{func:sec:convfunction:bkgFunc}
\end{multline*}
}
{
\begin{equation*}
  \mathcal{F}(\mllg; \vec{\mu}_G, \vec{\sigma}_G, s, \vec{\alpha}) \equiv \int_{-\infty}^{\infty} \rd{t} \mathcal{N}(\mllg - t; \vec{\mu}_G, \vec{\sigma}_G)\Theta(t;s)f(t;\vec{\alpha}),
    \label{func:sec:convfunction:bkgFunc}
\end{equation*}
}

where $t$ is the convolution variable, $\mathcal{N}(\mllg - t; \vec{\mu}_G, \vec{\sigma}_G)$ is a sum of one or two Gaussian pdfs with mean $\vec{\mu}_G = 0$ and standard deviation $\vec{\sigma}_G$,
$\Theta(t;s)$ is a step function that turns on at $t = s$, and $f(t;\vec{\alpha})$ is a smooth empirical function used to describe the slow falloff that depends on a set of parameters $\vec{\alpha}$.
Several families of falloff functions $f(t;\vec{\alpha})$ are considered: Bernstein polynomials, sums of power laws, sums of exponential functions, and Laurent series partial sums.
Each of these families consists of functions that are expressed as a sum of $n$ terms, where the values for the integer $n$ are determined in the process of selecting the fit functions, as described below. 
In addition to the convolution models, two Gaussian-variant models are also considered. In the modified Gaussian model, the width and the exponent in the exponential argument are allowed to vary linearly with \mllg. In the asymmetric generalized Gaussian model, the argument is not \mllg but
\begin{equation}
  y(\mllg;\alpha, \kappa,\xi)= -\frac{1}{\kappa}\ln\left[  1- \frac{\kappa(\mllg - \xi)}{\alpha}\right]
  \label{eq:aggargument}
\end{equation}
for constants $\alpha$, $\kappa$, and $\xi$, and the pdf is divided by $[\alpha-\kappa(\mllg-\xi)]$.

Starting with the full collection of convolution and Gaussian-variant models, a series of tests is applied to select a set of functions for each category that can describe the background well without biasing the extracted signal yield. 
The tests are as follows:

\begin{enumerate}

  \item To evaluate goodness of fit to data, a maximum likelihood fit is performed to the $\mllg$ distribution in each category, excluding the Higgs boson mass region. This region is not used to reduce bias. A $\chi^2$ test and a Kolmogorov--Smirnov (KS) test~\cite{Kolmogorov:1933,Smirnov:1948} are performed, and fit functions yielding a $p$-value of $p<0.10$ for the $\chi^2$ test or $p<0.20$ for the KS test are removed from the set of functions considered for a given category, as are models whose fit fails to reach an appropriate minimum. This test rejects a few functions that have insufficient freedom to describe the background shape, although most of the considered empirical functions pass these tests.
    
  \item In the ggF and VBF categories, even a small amount of background mismodeling can produce significant bias because of the large background. For this reason, the empirical functions are required to satisfy a \textit{spurious signal test}, which was introduced by ATLAS~\cite{ATLAS:2012yve,ATLAS:2014cnc}. This test assesses the background functions around the Higgs boson mass by using simulated background events. A fit is performed to the background-only simulated event samples, allowing for signal as well as background contributions in the fit model. The spurious signal $\zetas$ is defined as
\begin{equation}
  \abs{\zetas} = \text{max}(0, \abs{\nspurious} - 2\deltamc), 
\end{equation}
where $\deltamc$ is the uncertainty in the fit and $\nspurious$ is the fit signal yield when the fit is performed using the uncertainties expected in data. The quantity $\zetas$ represents the spurious signal yield after correcting for fluctuations in the simulated sample, which are small compared to those in data. To pass the test, $\zetas$ must be less than 20\% of the expected uncertainty in data, or less than 10\% of the expected SM signal yield. Like the goodness-of-fit test, the spurious signal test tends to reject functions with insufficient freedom to describe the background shapes.

\item To determine the maximum number of degrees of freedom, $n$, used in the falloff function $f(t;\vec{\alpha})$ for the convolution model families, an $\mathcal{F}$-test is performed~\cite{Fisher:1922saa}. This test provides a measure of whether there is a statistically significant improvement from the inclusion of additional terms. For the functions remaining in each family after the previous tests, a fit is performed to the data sidebands and $n$ is incremented until the $p$-value of the improvement is greater than 0.05, after which all higher-order functions are removed from the set.

\item A family of tests is used to select the final set of functions to avoid bias in the combined fit. For each function remaining in the set of background functions for each category, a fit is performed to the distribution in \mllg for the Higgs boson mass sideband region in data. Using the functions obtained from these fits as parent probability distributions, three sets of 1000 pseudo-experiments are generated: one with no signal added, one with the nominal signal model added to the background model, and one with five times the nominal signal added to the background model. The full set of functions is then used to fit each pseudo-experiment using the discrete profiling method, and the pull for each event sample is calculated as $(\mu-\widetilde{\mu})/\sigma$, where $\mu$ is the measured signal strength, $\sigma$ is the associated uncertainty, and $\widetilde{\mu}$ is the injected signal strength. For an unbiased signal extraction, one expects the pull distribution to be centered at zero with a standard deviation of one from statistical fluctuations. The absolute value of the mean pull for each of the three sets of pseudo-experiments is required to be less than 0.2 to ensure that the bias is small relative to the statistical uncertainty. If the mean is larger than 0.2, the functions with highest average bias when selected by the fit are removed.

\end{enumerate}

After applying these criteria, anywhere from 3 to 6 empirical background functions remain in each of the 13 categories and are used in the final statistical analysis.

\section{Systematic uncertainties\label{Sec-SystematicUncertainties}}

The simultaneous fit to the distribution in \mllg involves two main types of contributions: the smooth, nonresonant background, and the narrowly peaking resonant \HZG and \Hmm processes.
The uncertainty in the choice of nonresonant background shape is estimated using the discrete profiling method, which allows the fit to choose from the collection of functional forms validated as described in the previous section using a freely floating nuisance parameter.
This procedure effectively translates the uncertainty in the choice of background shape into a component of the statistical uncertainty.
In contrast, the signal and the resonant \Hmm background are described with models constructed using fits to simulated event samples. 
The sources of uncertainty in the resonant \HZG and \Hmm shapes are implemented in the statistical model as nuisance parameters that are floating but constrained by auxiliary measurements. 
The effects of these constrained nuisance parameters are referred to as systematic uncertainties.
These include both normalization uncertainties, which affect the resonant event yields and thus the DSCB scale, and shape uncertainties, which affect other parameters of the DSCB models.
The sizes of these effects are assessed by varying the relevant components of the simulated events samples.
The effects of normalization uncertainty sources are category- and process-dependent, providing freedom that captures both changes in the overall number of events and the migration of events between categories.
As such, the corresponding values below are expressed as ranges.

Several sources of uncertainty in the signal model arising from theory are considered. 
The uncertainty in fixed-order calculations used to determine the Higgs boson production cross sections, branching fractions, and \pt distributions is derived by varying the renormalization and factorization scales used in each calculation~\cite{deFlorian:2016spz,Bagnaschi:2011tu,Nason:2009ai,Luisoni:2013cuh,Hartanto:2015uka,Karlberg:2024zxx}.
The uncertainty in the fixed-order calculations allows the signal yields to vary by up to 11\% and contributes 9\% to the overall uncertainty in the signal strength, making it the largest source of systematic uncertainty.
An approximately 3\% modification to the signal cross section and a $<$1\MeV shift in the resonance peak are predicted to arise from the interference between the Higgs boson decay process and nonresonant \smzg production~\cite{Buccioni:2023qnt}, but an associated uncertainty is not included as it is small compared to other uncertainties.
The uncertainty in cross sections arising from input PDFs, quark masses \mq, and the strong coupling constant \alpS is assigned following Refs.~\cite{deFlorian:2016spz,Karlberg:2024zxx} and allows the signal yields to vary by up to 3\%.
The uncertainties in the underlying event and parton shower models are evaluated by varying the settings of these models, and allow signal yield variations of up to an average of 3\%, though in the \vhptmiss category specifically, variations of up to 41\% are allowed to model migration of events from the ggF categories.
The simulation statistical sample size is also considered and allows signal yield variations of up to 3\%.

Sources of uncertainty in the signal model arising from experimental inputs and effects are also assessed.
The uncertainty in the integrated luminosity collected by CMS amounts to 0.7\%, 1.4\%, and 1.3\% in Run~2, 2022, and 2023, respectively~\cite{CMS:2021xjt, CMS:LUM-20-001, CMS:LUM-22-001, CMS-DP-2024-068}, and are implemented in the statistical model by allowing signal yield variations of the same size.
The uncertainty in the distribution of pileup vertices is assessed by varying the weights used to correct simulation within the uncertainty of the measured total inelastic cross section, corresponding to allowed signal yield variations of up to 5\%.
The uncertainties in the signal yield due to the efficiency for reconstruction, identification, and triggering for electrons, muons, and photons are assessed using \Zee, \Zmm, and a combination of \Zee and \Zmmg events, respectively.
Additional uncertainty from trigger timing miscalibration is also assigned.
These efficiency uncertainty sources allow the signal yields to vary by up to 7\%.
The uncertainties in the electron, muon, and photon energy scales and resolutions are propagated both to the signal yields, where the effects are $<1\%$, and to the signal distribution in \mllg.
The uncertainties in the jet energy and \ptmiss scale and resolution allow up to a 4\% variation in the signal yields.
The uncertainties in b tagging for b, charm, and light-flavor or gluon jets are incorporated in the signal model and allow the signal yields to vary by up to 2\%.
Finally, the uncertainties in the photon identification variable, IDMVA, and the photon energy resolution are assessed using \Zmmg events and allow the signal yields to vary by up to 2\%.

Table~\ref{tab:systscale} shows the expected impact of the various analysis uncertainties on the signal strength.
The largest source of uncertainty is statistical.
The uncertainty in the choice of background shape functional form (discrete profiling) is evaluated by comparing the uncertainties when the function selection parameter is fixed to that obtained when it is floating.
The systematic uncertainty in the signal model is comparatively small, with theoretical uncertainty in fixed-order calculations making up the largest fraction.

\begin{table*}[tbp!]
\centering
\topcaption{Breakdown of sources of uncertainty. The statistical uncertainty is quantified by performing the analysis with the systematic uncertainty nuisance parameters frozen. The background shape functional form component of the statistical uncertainty is quantified by comparing a fit with the background functional form frozen to one where it is allowed to float. The sources of signal systematic uncertainty are quantified using the impact on the best fit signal strength when the auxiliary measurements are changed by $\pm$1 standard deviation.}
\label{tab:systscale}
\begin{tabular}{l c c c c}
\hline
Uncertainty source & $\Delta\mu$ \\
\hline
Statistical uncertainty &                      +0.49, -0.60 \\
\quad Background shape functional form (discrete profiling) &  +0.08, -0.36 \\
Systematic uncertainty &                       0.11 \\
\quad Fixed-order calculations &               0.09 \\
\quad Lepton and photon energy scale and resolution & 0.04 \\
\quad Photon identification variables &           0.04 \\
\quad Parton distribution functions, \alpS, and \mq & 0.03 \\
\quad Lepton and photon reconstruction, identification, and trigger efficiency & 0.02 \\
\quad Integrated luminosity and pileup &       0.01 \\
\quad Underlying event and parton shower &     0.01 \\
\quad Jet energy and \ptmiss, b tagging, simulation sample size &   $<$0.01 \\
\hline
\end{tabular}
\end{table*}

\section{Results\label{Sec-Results}}

\begin{figure*}[tb!]
  \centering
      \includegraphics[width=0.48\textwidth]{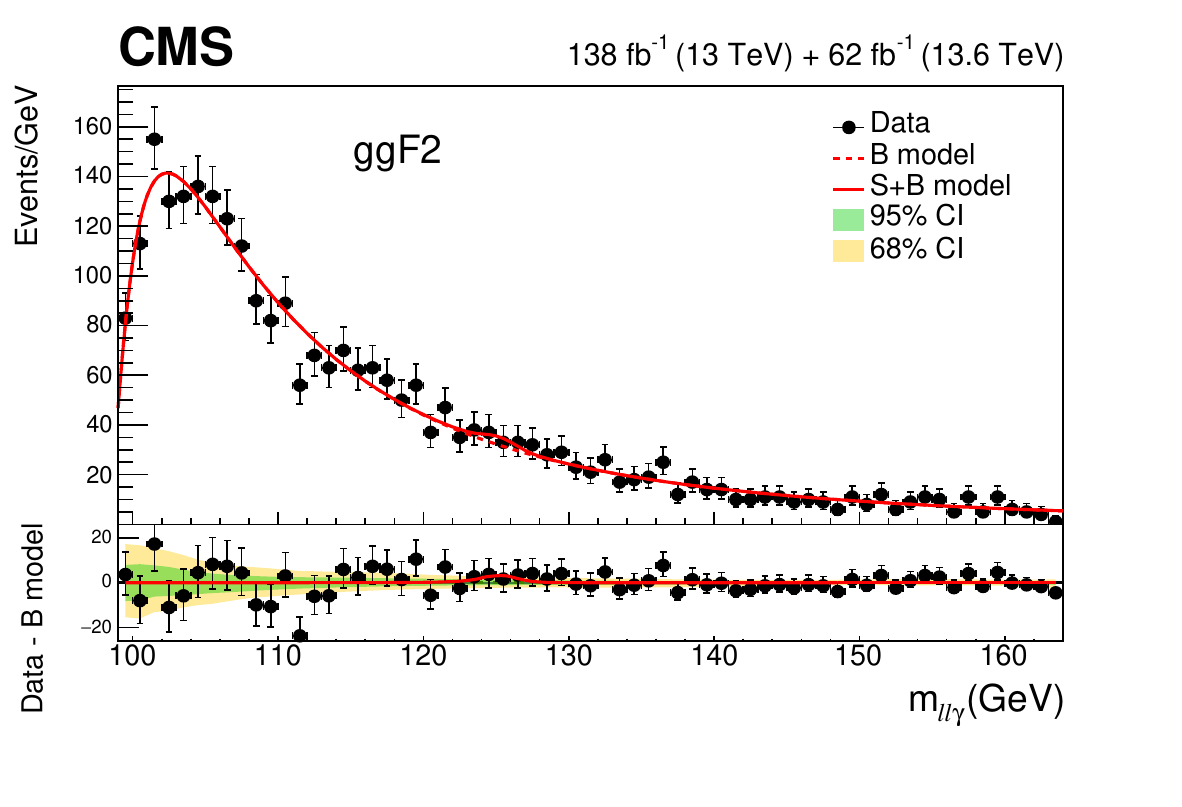}
      \includegraphics[width=0.48\textwidth]{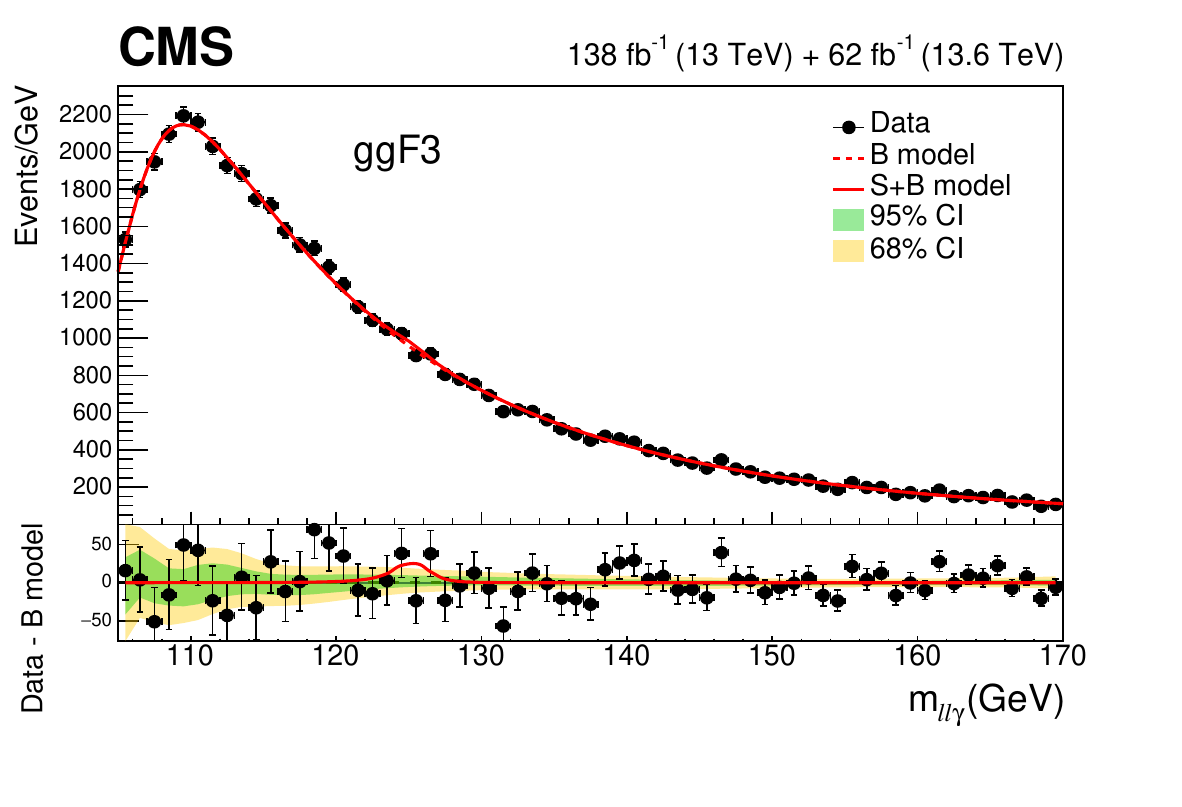}
  \centering
    \includegraphics[width=0.48\textwidth]{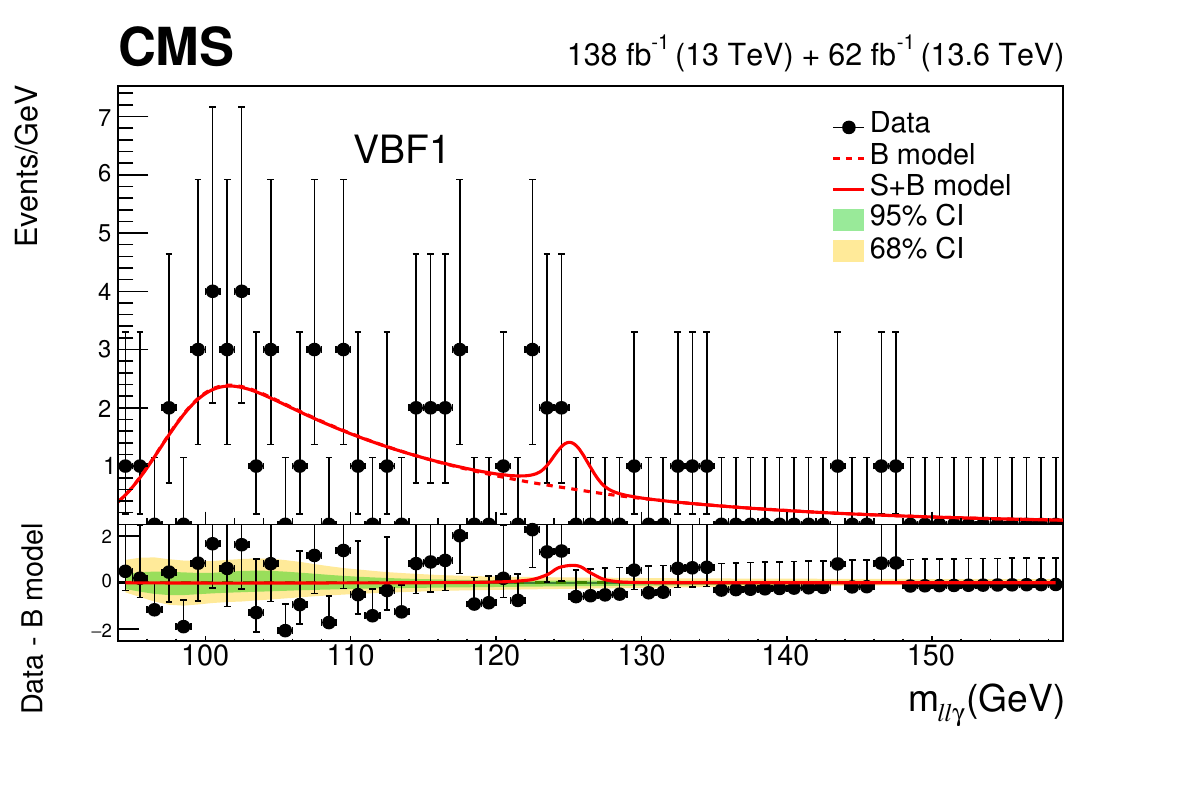}
    \includegraphics[width=0.48\textwidth]{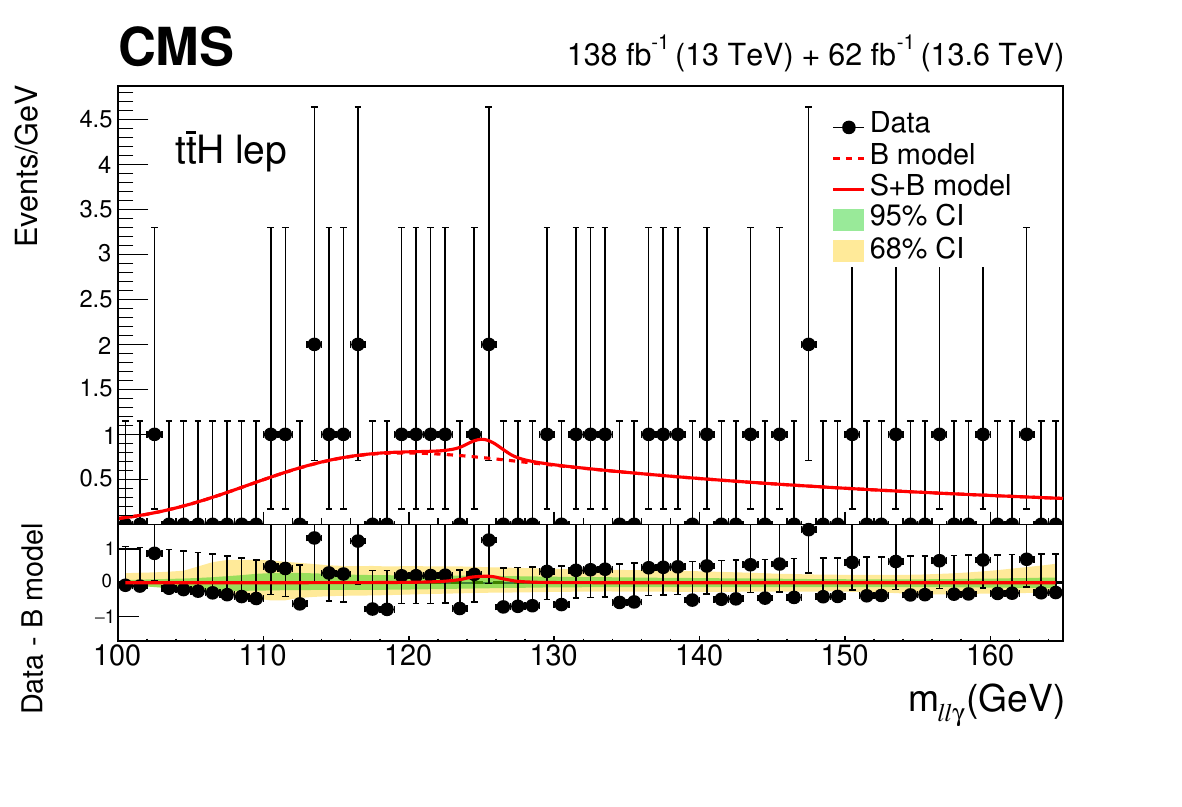}
    \caption{Distribution of data in \mllg for the (upper left) ggF2, (upper right) ggF3, (lower left) VBF1, and (lower right) \tthlep categories with the results of the simultaneous fit to all categories ($\text{S}+\text{B}$ model) superimposed. The $\text{B}$ model is the background-only component of the simultaneous fit. The \mllg range displayed for each category corresponds to the range over which the fit was performed.}
    \label{fig:sec-results-simultaneous-fits-ind-categories-fourexamples}
\end{figure*}

The maximum likelihood fit described in Section~\ref{Sec-FittingAndResults} is performed simultaneously in 13 categories.
Figure~\ref{fig:sec-results-simultaneous-fits-ind-categories-fourexamples} shows the results of the simultaneous fit in the \mllg distributions for four of these categories. 
The fit shows good agreement between the background model and the data in the \mllg regions below and above the Higgs boson mass region in all 13 categories, and no systematic trends are observed in the residual distributions shown in the lower panels.

\begin{figure}[tbp]
    \centering
    \includegraphics[width=0.49\textwidth]{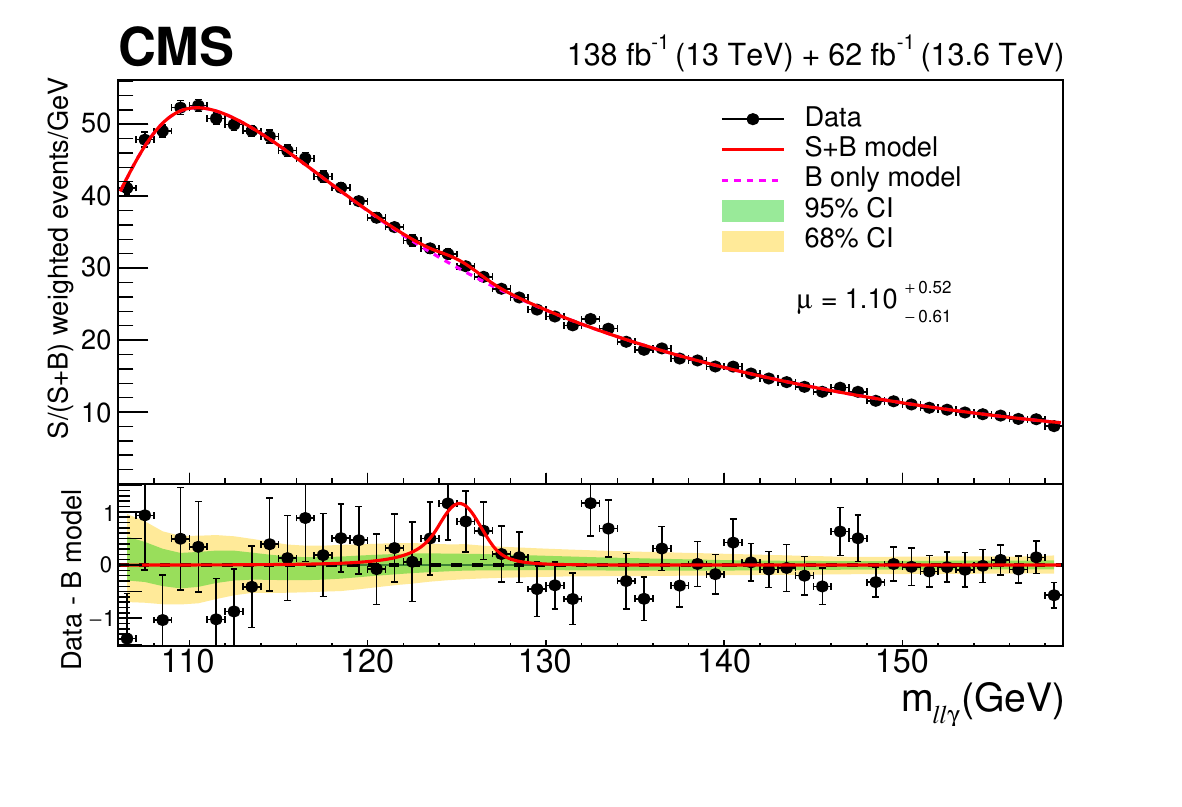}
    \caption{Weighted distribution of data in $\mllg$ and the signal-plus-background model obtained from the simultaneous fit.
    The data and model comprise all 13 categories, weighted by the factor $S/(S+B)$.}
    \label{fig:sec-results-combined}
\end{figure}

Figure~\ref{fig:sec-results-combined} shows the combined result from the simultaneous fit in a single histogram obtained by weighting each category by $S/(S+B)$, where $S$ is the expected signal yield in the category and $B$ is the integrated background in the Higgs boson mass region.

\begin{figure}[tb!]
    \centering
    \includegraphics[width=0.49\textwidth]{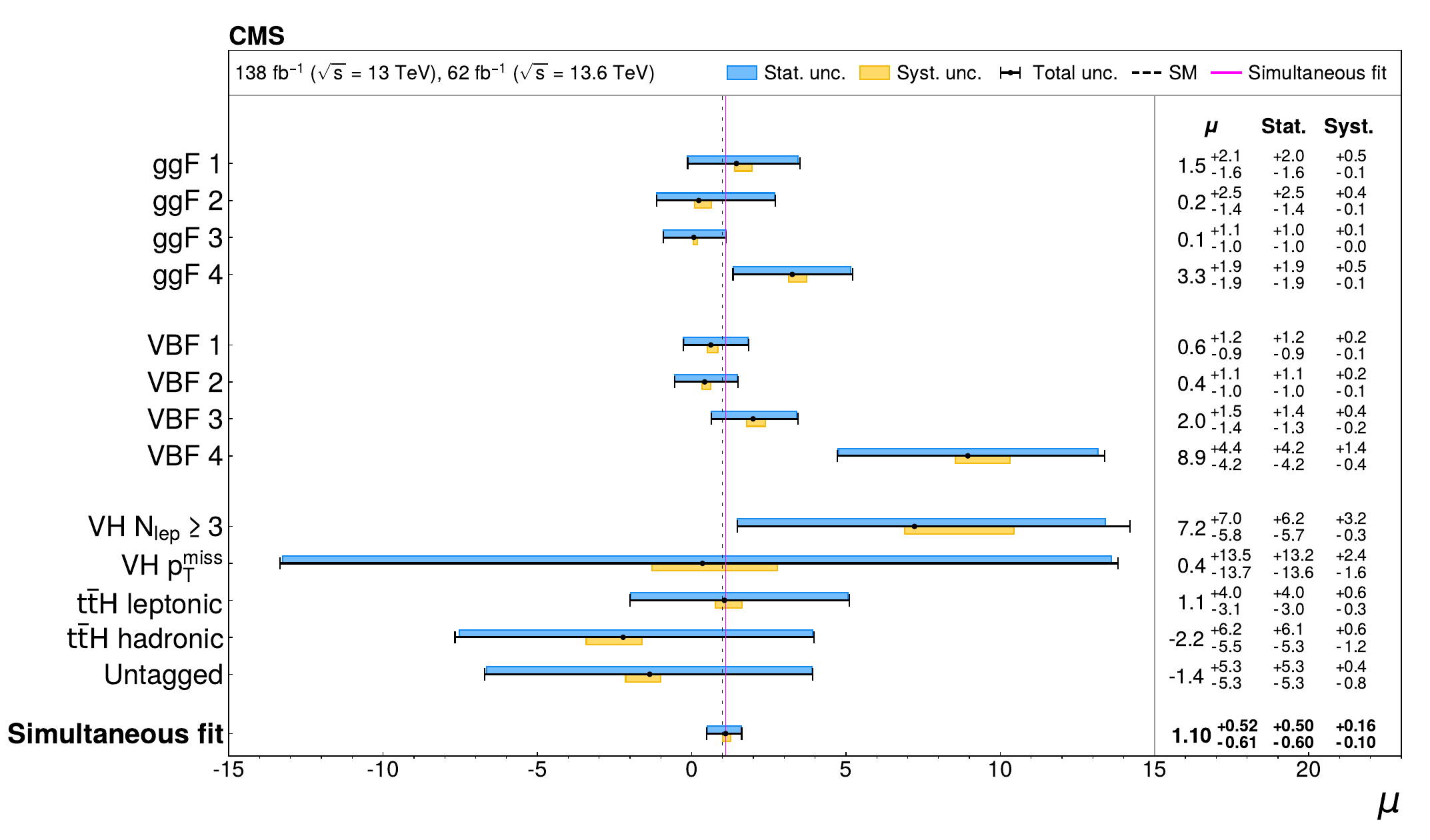}
    \caption{Signal strengths ($\mu$) obtained from separate fits to the \mllg distribution in each category and, as the last entry in the figure, from the simultaneous fit to all categories.
    The reported systematic uncertainty is the difference in quadrature between the uncertainty calculated with the signal nuisance parameters frozen and that with the signal nuisance parameters left free to vary.
    As a result, the reported systematic uncertainty can be asymmetric, even if the nuisance parameter constraints in the likelihood are symmetric.}
    \label{fig:sec-results-signal-strengths}
\end{figure}

Figure~\ref{fig:sec-results-signal-strengths} lists the signal strengths $\mu$ obtained in each category, as well as the result from the simultaneous fit. 
As expected, the VBF and ggF event categories have the smallest uncertainties.
The $p$-value of the simultaneous fit is $p=0.75$, indicating a high degree of compatibility across the categories.
For $\mH = 125.38\GeV$, the observed (expected) signal strength obtained from the simultaneous fit is $\mu = 1.10 ^{+0.52}_{-0.61}\ (1.00^{+0.49}_{-0.46})$,
consistent with the SM prediction. The observed
(expected) significance of the excess is $1.9$ (2.3) standard deviations.
Separating the fit uncertainty into systematic and statistical components,
we obtain $\mu = 1.10 ^{+0.49}_{-0.60}\stat^{+0.18}_{-0.07}\syst$. 

\section{Summary\label{Sec-Summary}}

This paper presents a search for the rare Higgs boson decay $\HZG$, where $\Zll$ and $\Pell = \Pe$ or $\PGm$. The standard model (SM) branching fraction, not including $\mathcal{B}(\Zll)$, is predicted to be $\mathcal{B}(\HZG) = (1.5 \pm 0.1) \times 10^{-3}$~\cite{Djouadi:1997yw,deFlorian:2016spz}.
The search is performed using a sample of proton-proton ($\Pp\Pp$) collision data at the center-of-mass energies of $13$ and $13.6\TeV$, recorded by the CMS experiment at the LHC between 2016 and 2023. The sample corresponds to a total integrated luminosity of 200\fbinv.

Measurement of the $\HZG$ signal is complicated by large nonresonant backgrounds from other SM processes
that produce the same reconstructed objects in the detector. In most production channels,
the dominant background arises from $\Pp\Pp\to\PZ\PGg$ events,
with an additional significant contribution from $\Pp\Pp\to\PZ$ events in which the photon candidate arises from misreconstruction, nonprompt decays, jet fragmentation, multiparton interactions, or separate $\Pp\Pp$ interaction vertices.

The analysis separately considers and optimizes sensitivity to Higgs boson production via gluon-gluon fusion (ggF), vector boson fusion (VBF), and associated-production processes. 
This analysis uses four ggF categories, four VBF categories, four associated-production categories, and one untagged category. 
The signal is then extracted from a simultaneous maximum likelihood fit to the invariant mass distributions of the $\llg$ system in the various event categories, employing the discrete profiling method and an extensive series of validation tests.

The simultaneous maximum likelihood fit to all 13 event categories yields
an observed (expected) signal strength $\mu$, defined as the ratio of the measured product of the cross section and decay branching fraction $\sigma(\ppH)\mathcal{B}(\HZG)$ to the corresponding SM prediction, of $\mu = 1.10 ^{+0.52}_{-0.61}\ (1.00^{+0.49}_{-0.46})$ for a Higgs boson mass of $125.38\GeV$.
The $p$-value associated with the compatibility of the 13 separate event categories is $0.75$, and the signal has an observed (expected) significance of 1.9 (2.3) standard deviations.

\begin{acknowledgments}
We congratulate our colleagues in the CERN accelerator departments for the excellent performance of the LHC and thank the technical and administrative staffs at CERN and at other CMS institutes for their contributions to the success of the CMS effort. In addition, we gratefully acknowledge the computing centers and personnel of the Worldwide LHC Computing Grid and other centers for delivering so effectively the computing infrastructure essential to our analyses. Finally, we acknowledge the enduring support for the construction and operation of the LHC, the CMS detector, and the supporting computing infrastructure provided by the following funding agencies: SC (Armenia), BMFWF and FWF (Austria); FNRS and FWO (Belgium); CNPq, CAPES, FAPERJ, FAPERGS, and FAPESP (Brazil); MES and BNSF (Bulgaria); CERN; CAS, MoST, and NSFC (China); MINCIENCIAS (Colombia); MSES and CSF (Croatia); RIF (Cyprus); SENESCYT (Ecuador); ERC PRG and PSG, TARISTU24-TK10 and MoER TK202 (Estonia); Academy of Finland, MEC, and HIP (Finland); CEA and CNRS/IN2P3 (France); SRNSF (Georgia); BMFTR, DFG, and HGF (Germany); GSRI (Greece); MATE and NKFIH (Hungary); DAE and DST (India); IPM (Iran); SFI (Ireland); INFN (Italy); MSIT and NRF (Republic of Korea); MES (Latvia); LMTLT (Lithuania); MOE and UM (Malaysia); BUAP, CINVESTAV, CONACYT, LNS, SEP, and UASLP-FAI (Mexico); MOS (Montenegro); MBIE (New Zealand); PAEC (Pakistan); MSHE, NSC, and NAWA (Poland); FCT (Portugal);  MESTD (Serbia); MICIU/AEI and PCTI (Spain); MOSTR (Sri Lanka); Swiss Funding Agencies (Switzerland); MST (Taipei); MHESI (Thailand); TUBITAK and TENMAK (T\"{u}rkiye); NASU (Ukraine); STFC (United Kingdom); DOE and NSF (USA).

\begin{sloppypar}
\setlength\emergencystretch{\hsize}
\hyphenation{Rachada-pisek} Individuals have received support from the Marie-Curie program and the European Research Council and Horizon 2020 Grant, contract Nos.\ 675440, 724704, 752730, 758316, 765710, 824093, 101115353, 101002207, 101001205, and COST Action CA16108 (European Union); the Leventis Foundation; the Alfred P.\ Sloan Foundation; the Alexander von Humboldt Foundation; the Science Committee, project no. 22rl-037 (Armenia); the Fonds pour la Formation \`a la Recherche dans l'Industrie et dans l'Agriculture (FRIA) and Fonds voor Wetenschappelijk Onderzoek contract No. 1228724N (Belgium); the Beijing Municipal Science \& Technology Commission, No. Z191100007219010, the Fundamental Research Funds for the Central Universities, the Ministry of Science and Technology of China under Grant No. 2023YFA1605804, the Natural Science Foundation of China under Grant No. 12535004, and USTC Research Funds of the Double First-Class Initiative No.\ YD2030002017 (China); the Ministry of Education, Youth and Sports (MEYS) of the Czech Republic; the Shota Rustaveli National Science Foundation (Georgia); the Deutsche Forschungsgemeinschaft (DFG), among others, under Germany's Excellence Strategy -- EXC 2121 ``Quantum Universe" -- 390833306, and under project number 400140256 - GRK2497; the Hellenic Foundation for Research and Innovation (HFRI), Project Number 2288 (Greece); the Hungarian Academy of Sciences, the New National Excellence Program - \'UNKP, the NKFIH research grants K 131991, K 138136, K 143460, K 143477, K 147557, K 146913, K 146914, K 147048, TKP2021-NKTA-64, and 2025-1.1.5-NEMZ\_KI-2025-00004, and MATE KKP and KKPCs Research Excellence and Flagship Research Groups grants (Hungary); the Council of Science and Industrial Research, India; ICSC -- National Research Center for High Performance Computing, Big Data and Quantum Computing, FAIR -- Future Artificial Intelligence Research, and CUP I53D23001070006 (Mission 4 Component 1), funded by the NextGenerationEU program, the Italian Ministry of University and Research (MUR) under Bando PRIN 2022 -- CUP I53C24002390006, PRIN PRIMULA 2022RBYK7T (Italy); the Latvian Council of Science; the Ministry of Science and Higher Education, project no. 2022/WK/14, and the National Science Center, contracts Opus 2021/41/B/ST2/01369, 2021/43/B/ST2/01552, 2023/49/B/ST2/03273, and the NAWA contract BPN/PPO/2021/1/00011 (Poland); the Funda\c{c}\~ao para a Ci\^encia e a Tecnologia (Portugal); the National Priorities Research Program by Qatar National Research Fund; MICIU/AEI/10.13039/501100011033, ERDF/EU, ``European Union NextGenerationEU/PRTR", projects PID2022-142604OB-C21, PID2022-139519OB-C21, PID2023-147706NB-I00, PID2023-148896NB-I00, PID2023-146983NB-I00, PID2023-147115NB-I00, PID2023-148418NB-C41, PID2023-148418NB-C42, PID2023-148418NB-C43, PID2023-148418NB-C44, PID2024-158190NB-C22, RYC2021-033305-I, RYC2024-048719-I, CNS2023-144781, CNS2024-154769 and Plan de Ciencia, Tecnolog{\'i}a e Innovaci{\'o}n de Asturias, Spain; the Chulalongkorn Academic into Its 2nd Century Project Advancement Project, the National Science, Research and Innovation Fund program IND\_FF\_68\_369\_2300\_097, and the Program Management Unit for Human Resources \& Institutional Development, Research and Innovation, grant B39G680009 (Thailand); the Eric \& Wendy Schmidt Fund for Strategic Innovation through the CERN Next Generation Triggers project under grant agreement number SIF-2023-004; the Kavli Foundation; the Nvidia Corporation; the SuperMicro Corporation; the Welch Foundation, contract C-1845; and the Weston Havens Foundation (USA).
\end{sloppypar}
\end{acknowledgments}\section*{Data availability} Release and preservation of data used by the CMS Collaboration as the basis for publications is guided by the  \href{https://doi.org/10.7483/OPENDATA.CMS.1BNU.8V1W}{CMS data preservation, re-use and open access policy}.
\bibliography{auto_generated} 
\cleardoublepage \appendix\section{The CMS Collaboration \label{app:collab}}\begin{sloppypar}\hyphenpenalty=5000\widowpenalty=500\clubpenalty=5000\cmsinstitute{Yerevan Physics Institute, Yerevan, Armenia}
{\tolerance=6000
A.~Gevorgyan\cmsorcid{0000-0003-2751-9489}, A.~Hayrapetyan, V.~Makarenko\cmsorcid{0000-0002-8406-8605}, A.~Tumasyan\cmsAuthorMark{1}\cmsorcid{0009-0000-0684-6742}
\par}
\cmsinstitute{Marietta Blau Institute for Particle Physics, Vienna, Austria}
{\tolerance=6000
W.~Adam\cmsorcid{0000-0001-9099-4341}, L.~Benato\cmsorcid{0000-0001-5135-7489}, T.~Bergauer\cmsorcid{0000-0002-5786-0293}, M.~Dragicevic\cmsorcid{0000-0003-1967-6783}, S.~Gundacker\cmsorcid{0000-0003-2087-3266}, A.K.~Guven\cmsorcid{0009-0004-5670-5138}, P.S.~Hussain\cmsorcid{0000-0002-4825-5278}, M.~Jeitler\cmsAuthorMark{2}\cmsorcid{0000-0002-5141-9560}, N.~Krammer\cmsorcid{0000-0002-0548-0985}, A.~Li\cmsorcid{0000-0002-4547-116X}, D.~Liko\cmsorcid{0000-0002-3380-473X}, M.~Matthewman, A.~Pfeiffer\cmsorcid{0000-0001-5328-448X}, J.~Schieck\cmsAuthorMark{2}\cmsorcid{0000-0002-1058-8093}, R.~Sch\"{o}fbeck\cmsAuthorMark{2}\cmsorcid{0000-0002-2332-8784}, M.~Shooshtari\cmsorcid{0009-0004-8882-4887}, M.~Sonawane\cmsorcid{0000-0003-0510-7010}, N.~Van~Den~Bossche\cmsorcid{0000-0003-2973-4991}, W.~Waltenberger\cmsorcid{0000-0002-6215-7228}, C.E.~Wulz\cmsAuthorMark{2}\cmsorcid{0000-0001-9226-5812}
% ...
\par}
$^{1}$Also at Yerevan State University, Yerevan, Armenia\\
$^{2}$Also at Technische Universit\"{a}t Wien, Vienna, Austria\\
% ...
\end{sloppypar}
\end{document}